\documentclass[aps, prc, twocolumn, superscriptaddress, showpacs, showkeys, nofootinbib]{revtex4-1}

\usepackage{amsmath, amsfonts, amssymb, bm}
\usepackage{graphicx}
\usepackage{color}
\usepackage{subfigure}
\usepackage{multirow}
\usepackage{textcomp}
\usepackage{slashed}

\usepackage[mathscr,scaled=1.15]{urwchancal}
\DeclareFontFamily{OT1}{pzc}{}
\DeclareFontShape{OT1}{pzc}{m}{it}%
{<-> s * [1.15] pzcmi7t}{}
\DeclareMathAlphabet{\mathpzc}{OT1}{pzc}{m}{it}

\definecolor{purple}{rgb}{0.5,0,0.5}
\definecolor{blue}{rgb}{0.0,0,0.9}
\definecolor{prdblue}{rgb}{0.133,0.118,0.498}
\usepackage[colorlinks=true, pdfstartview=FitV, linkcolor=prdblue, citecolor= prdblue, urlcolor=prdblue]{hyperref}

\begin{document}


\title{Nucleon-to-Roper electromagnetic transition form factors at large-$Q^2$}



\author{Chen Chen}
\affiliation{Instituto de F\'isica Te\'orica, Universidade Estadual Paulista, Rua Dr.~Bento Teobaldo Ferraz, 271,
01140-070 S\~ao Paulo, SP, Brazil}

\author{Ya Lu}
\affiliation{Department of Physics, Nanjing University, Nanjing, Jiangsu 210093, China}

\author{Daniele Binosi}
\affiliation{European Centre for Theoretical Studies in Nuclear Physics
and Related Areas (ECT$^\ast$) and Fondazione Bruno Kessler\\ Villa Tambosi, Strada delle Tabarelle 286, I-38123 Villazzano (TN) Italy}

\author{Craig D. Roberts}
\affiliation{Physics Division, Argonne National Laboratory, Argonne, Illinois 60439, USA}

\author{Jose Rodr\'{\i}guez-Quintero}
\affiliation{Department of Integrated Sciences, University of Huelva, E-21071 Huelva, Spain}

\author{Jorge Segovia}
\affiliation{Departamento de Sistemas F\'{\i}sicos, Qu\'{\i}micos y Naturales,
Universidad Pablo de Olavide, E-41013 Sevilla, Spain}




\date{20 November 2018}

\begin{abstract}
High-precision nucleon-resonance electroproduction data on a large kinematic domain of energy and momentum transfer have proven crucial in revealing novel features of strong interactions within the Standard Model and unfolding structural details of baryon excited states.  Thus, in anticipation of new data reaching to unprecedented photon virtuality, we employ a quark-diquark approximation to the three valence-quark bound-state problem to compute $\gamma^\ast p \to R^+$ and $\gamma^\ast n \to R^0$ transition form factors on $Q^2/m_N^2 \in [0,12]$, where $m_N$ is the nucleon mass.  Having simultaneously analysed both charged and neutral channels, we also provide a quark-flavour separation of the transition form factors.  The results should be useful in planning new-generation experiments.
\end{abstract}

\maketitle


\section{Introduction}
The task of mapping and explaining the spectrum of baryons and the structure of these states is a longstanding challenge, which is likely to stand for another decade or more \cite{Isgur:2000ad}.  The ground-state neutron and proton (nucleons) are certainly bound-states seeded by three valence-quarks: $udd$ and $uud$, respectively.  However, the natures of the nucleons' first excited states -- $N(1440)\,1/2^+$, $N(1535)\,1/2^-$ -- are less certain.  The $N(1440)\,1/2^+$ ``Roper resonance'' was discovered in 1963 \cite{Roper:1964zza, BAREYRE1964137, AUVIL196476, PhysRevLett.13.555, PhysRev.138.B190}, but it was immediately a source of puzzlement because, \emph{e.g}.\ a wide array of constituent-quark potential models produce a spectrum in which the second positive-parity state in the baryon spectrum lies above the first negative-parity state \cite{Capstick:2000qj, Crede:2013sze, Giannini:2015zia}.

In connection with the Roper, the last twenty years have seen the acquisition and analysis of a vast amount of high-precision proton-target exclusive electroproduction data with single- and double-pion final states on a large kinematic domain of energy and momentum-transfer; development of a sophisticated dynamical reaction theory capable of simultaneously describing all partial waves extracted from available, reliable data; and formulation and wide-ranging application of a Poincar\'e covariant approach to the continuum bound state problem in relativistic quantum field theory.  Following these efforts, it is now widely accepted that the Roper is, at heart, the first radial excitation of the nucleon, consisting of a well-defined dressed-quark core that is augmented by a meson cloud, which both reduces the Roper's core mass by approximately 20\% and contributes materially to the electroproduction transition form factors at low-$Q^2$ \cite{Golli:2017nid, Burkert:2017djo}.


The high-$Q^2$ electroproduction data were crucial to reaching this understanding of the Roper: the short-wavelength probe pierces the long-wavelength screen generated by meson-baryon final state interactions (MB\,FSIs) and thereby reveals the dressed-quark core that forms the core of a true resonance.  Regarding the charged-Roper, accurate $\gamma^\ast p$ electroproduction data is now available on the kinematic range $W \leq 2\,$GeV and $Q^2\leq 4.5\,$GeV$^2$ \cite{Aznauryan:2004jd, Aznauryan:2008pe, Dugger:2009pn, Aznauryan:2011qj, Aznauryan:2009mx, Mokeev:2012vsa, Mokeev:2015lda,Aznauryan:2012ec, Aznauryan:2012ba,  Park:2014yea, Isupov:2017lnd, Fedotov:2018oan, Burkert:2016dxc, Burkert:2018oyl}.  In the near future, with a new era of experiments beginning at the upgraded Thomas Jefferson National Accelerator Facility (JLab\,12), there is the potential to match this with high-precision electroproduction data off a (bound-)\,neutron target and therewith chart neutral-Roper properties.  This will provide an important additional test for theory because a successful, unified explanation of nucleon, $\Delta$-baryon, and charged and neutral Roper elastic and transition form factors will sharpen the contemporary picture of the Roper resonance as, primarily, the nucleons' first radial excitation.

Poincar\'e-covariant continuum analyses of the three valence-quark bound-state problems associated with the nucleon, $\Delta$-baryon and Roper resonance are presented in Refs.\,\cite{Segovia:2014aza, Roberts:2015dea, Segovia:2015hra, Segovia:2016zyc, Chen:2017pse, Mezrag:2017znp, Roberts:2018hpf}: the same framework is used for all systems.  Importantly, following an appreciation of its importance for the ground state nucleons \cite{Cates:2011pz}, Ref.\,\cite{Segovia:2016zyc} computes a quark-flavour separation of the nucleon-to-Roper transition.  Missing therein, however, are predictions for the charged and neutral Roper elastic form factors, and the $\gamma^\ast n \to R^0$ transition form factors, which were actually used in order to compute the flavour-separated results.  Given that JLab\,12 will deliver results for the $Q^2$-dependence of $R^{0,+}$ electrocouplings on a hitherto unexplored domain \cite{Mokeev:2018zxt, Carman:2018fsn, Cole:2018faq}, reaching to $Q^2\approx 12\,m_N^2$, where $m_N$ is the nucleon mass, herein we report calculations of all Roper-related transition form factors, including those not discussed in Refs.\cite{Segovia:2015hra, Segovia:2016zyc}, on the complete $Q^2$-domain that is expected to be mapped by the planned and foreseen experiments.

Section\,\ref{secNRStructure} provides a succinct explanation of the nucleon and Roper mass and wave function calculations that provide the input to our computation of the related transition form factors.  The electromagnetic vertex we use in these calculations is detailed in Sec.\,\ref{secEMcurrent}.
Section~\ref{secEMelastic} describes the Roper resonance elastic form factors, comparing them with those of the nucleon ground states and presenting a breakdown into contributions from different correlation sectors.
%
%
Whilst measurement of such elastic form factors presents many challenges,
their theoretical analysis reveals valuable structural information; and they are necessary to fix the canonical normalisation of the bound-state amplitudes.
The $\gamma^\ast p \to R^+$, $\gamma^\ast n \to R^0$ transition form factors on $Q^2 \lesssim 6\,m_N^2$ are analysed in Sec.\,\ref{secEMtransition} and compared with available data.  Projections onto the domain made accessible with JLab\,12, \emph{viz}.\ $6 \lesssim Q^2/{\rm GeV}^2\lesssim 12$, obtained using a novel numerical technique, are described in Sec.\,\ref{SecLargeQ2}; and  flavour-separated transition form factors on the entire domain $Q^2 \lesssim 12\,m_N^2$ are presented in Sec.\,\ref{secFlavourSeparated}.  Section~\ref{secEpilogue} provides a summary and perspective.


\section{Nucleon and Roper Structure}
\label{secNRStructure}
\begin{figure}[t]
\centerline{%
\includegraphics[clip, width=0.45\textwidth]{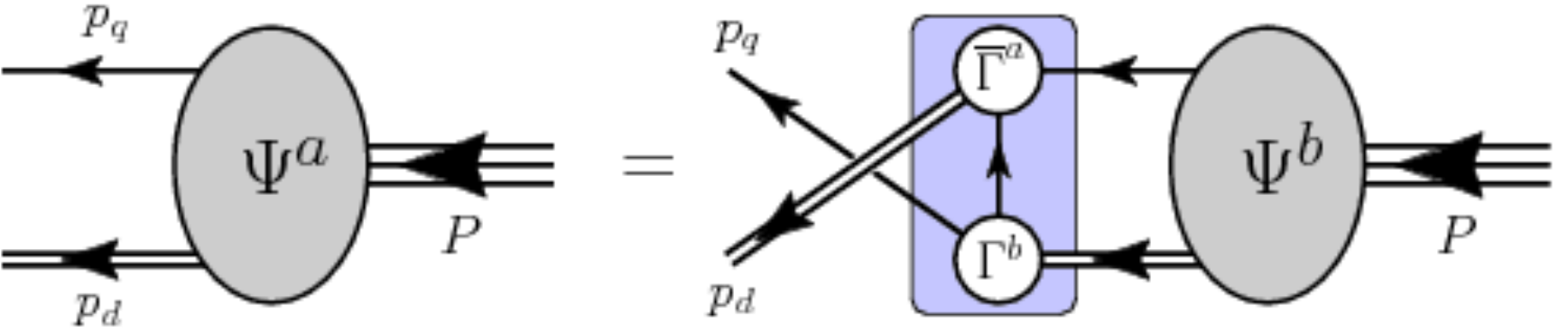}}
\caption{\label{figFaddeev}
Poincar\'e covariant Faddeev equation: a linear integral equation for the matrix-valued function $\Psi$, being the Faddeev amplitude for a baryon of total momentum $P= p_q + p_d$, which expresses the relative momentum correlation between the dressed-quarks and -diquarks within the baryon.  The shaded rectangle demarcates the kernel of the Faddeev equation: \emph{single line}, dressed-quark propagator; $\Gamma$,  diquark correlation amplitude; and \emph{double line}, diquark propagator. }
\end{figure}

We treat baryons as a continuum three--valence-body bound-state problem using the Poincar\'e-covariant Faddeev equation introduced in Refs.\,\cite{Cahill:1988dx, Burden:1988dt, Cahill:1988zi, Reinhardt:1989rw, Efimov:1990uz}.  The Faddeev equation sums all possible exchanges and interactions that can take place between the three dressed-quarks that express the baryon's valence-quark content; and used with a realistic quark-quark interaction \cite{Binosi:2014aea, Binosi:2016nme, Rodriguez-Quintero:2018wma}, it predicts the appearance of soft (nonpointlike) diquark correlations within baryons, whose characteristics are determined by dynamical chiral symmetry breaking (DCSB) \cite{Segovia:2015ufa}.  Consequently, the problem of determining the properties of the baryon's dressed-quark core is transformed into that of solving the linear, homogeneous matrix equation depicted in Fig.\,\ref{figFaddeev}.

Computation of the mass and structure of the nucleon and its first radial excitation is described in Ref.\,\cite{Segovia:2015hra}.  That analysis began by solving the Faddeev equation to obtain the masses and Poincar\'e-covariant wave functions for these systems, taking each element of the equation to be as specified in \cite{Segovia:2014aza}, which provides a successful description of the properties of the nucleon and $\Delta$-baryon.  With those inputs, the masses are (in GeV):
\begin{subequations}
\label{eqMasses}
\begin{align}
m_{{\rm nucleon}\,(N)} & = 1.18\,,\;\\
m_{{\rm nucleon-excited}\,(R)} & = 1.73\,.
\end{align}
\end{subequations}
The analysis predicts that the nucleon and Roper quark-cores are constituted only from isoscalar-scalar and isovector-pseudovector diquarks, with correlations in the pseudoscalar and vector channels having negligible impact on their properties, an outcome confirmed by other studies \cite{Eichmann:2016hgl, Lu:2017cln, Chen:2017pse}.

The masses in Eq.\,\eqref{eqMasses} correspond to the locations of the two lowest-magnitude $I=1/2$, $J^P=1/2^+$ poles in the three dressed-quark scattering problem.\footnote{As evident in Fig.\,\ref{figFaddeev}, MB\,FSIs are omitted from the scattering kernel.  Hence, these poles are real-valued.  This is discussed further in the following.}
The associated residues are the canonically-normalized Faddeev wave functions, which depend upon $(\ell^2,\ell \cdot P)$, where $\ell$ is the quark-diquark relative momentum and $P$ is the baryon's total momentum.  It is now useful to consider the zeroth Chebyshev moment of all $S$-wave components in that wave function, \emph{i.e}.\ projections of the form
\begin{equation}
{\mathpzc W}(\ell^2;P^2) = \frac{2}{\pi} \int_{-1}^1 \! du\,\sqrt{1-u^2}\,
{\mathpzc W}(\ell^2,u; P^2)\,,
\end{equation}
where $u=\ell\cdot P/\sqrt{\ell^2 P^2}$.
For the ground-state nucleon, these projections are either positive- or negative-definite; but for the first excited state, each exhibits a single zero.  (See, \emph{e.g}.\ Ref.\,\cite{Segovia:2015hra}, Fig.\,2, or Ref.\,\cite{Chen:2017pse}, Fig.\,4.)   Drawing upon experience with excited-state mesons studied via the Bethe-Salpeter equation \cite{Holl:2004fr, Qin:2011xq}, the appearance of a single zero in $S$-wave components of the Faddeev wave function associated with the first excited state in the scattering problem indicates that this state is a radial excitation of the quark-diquark system.  Notably, one may associate a four-vector length-scale of $1/[0.4 {\rm GeV}]\approx 0.5\,$fm with the location of this zero.

We do not consider the complementary case of radial excitation of the diquark itself without an excitation of the quark-diquark correlation.  As explained elsewhere \cite{Roberts:2011cf}, orthogonality of ground- and excited-state diquark correlations is likely to suppress any such admixture.  Notwithstanding that, it is probable that the possibility can most effectively be explored in the future via appropriate light-front projection \cite{Mezrag:2017znp} of solutions to the truly three-body bound-state equation, employed, \emph{e.g}.\ in Refs.\,\cite{Eichmann:2009qa, Qin:2018dqp}.   Here we only remark that the comparisons with data presented below support our picture of the Roper resonance as primarily a radial excitation in the quark-diquark relative momentum correlation.  

The structure of the $N(1710)\,1/2^+$ is less clear.  In quark models, the profile of its wave function is  sensitive to the formulation employed, \emph{e.g}.\ it can be Roper-like, with two peaks skewed relative to those in the kindred Roper wave function \cite{Capstick:1992xn, Melde:2008yr}, in which case it may be a candidate for the system which is predominantly quark-plus-radially-excited-diquark; or it can have three peaks, located on the same trajectory as the two in the related Roper wave function \cite{Santopinto:2012nq, deTeramond:2014asa}, \emph{viz}.\ the second radial excitation of the quark-plus-diquark system.  A third possibility, realised in some dynamical coupled channels (DCC) calculations \cite{Suzuki:2009nj}, sees the Roper and $N(1710)\,1/2^+$ as both derived from the same quark core state.  Given that $N(1710)\,1/2^+$ electroproduction data exist on $Q^2\lesssim 4\,m_N^2$ \cite{Park:2014yea} and that each helicity amplitude appears to be of unique sign, unlike those for the Roper \cite{Aznauryan:2009mx, Mokeev:2012vsa, Mokeev:2015lda}, in future it is worth testing these possibilities by exploring the solution space of our Poincar\'e-covariant Faddeev equation and using the results to compute the transition form factors.

Let us return now to consider the masses in Eq.\,\eqref{eqMasses}.  As elucidated in Ref.\,\cite{Suzuki:2009nj}, the empirical values of the pole locations for the first two states in the nucleon channel are: $0.939\,$GeV for the nucleon; and two poles for the Roper, $1.357 - i \,0.076$, $1.364 - i \, 0.105\,$GeV.  At first glance, these values appear unrelated to those in Eq.\,\eqref{eqMasses}.  However, deeper analysis reveals \cite{Eichmann:2008ae, Eichmann:2008ef} that the kernel in Fig.\,\ref{figFaddeev} omits all those resonant contributions which may be associated with the MB\,FSIs resummed in DCC models  \cite{JuliaDiaz:2007kz, Suzuki:2009nj, Kamano:2010ud, Ronchen:2012eg, Kamano:2013iva, Doring:2014qaa, Kamano:2018sfb} in order to transform a bare-baryon into the observed state.  The  Faddeev equation that produced the results in Eq.\,\eqref{eqMasses} should therefore be understood to describe not the completely-dressed and hence observable object, but rather this bare system, which we described above as the \emph{dressed-quark core} of the bound-state.

Clothing the nucleon's dressed-quark core by including resonant contributions to the kernel produces a physical nucleon whose mass is $\approx 0.2$\,GeV lower than that of the core \cite{Ishii:1998tw, Hecht:2002ej}.  Similarly, clothing the $\Delta$-baryon's core lowers its mass by $\approx 0.16\,$GeV \cite{JuliaDiaz:2007kz}.   It is therefore no coincidence that (in GeV) $1.18-0.2 = 0.98\approx 0.94$, \emph{i.e}.\ the nucleon mass in Eq.\,\eqref{eqMasses} is 0.2\,GeV greater than the empirical value.  A successful body of work \cite{Segovia:2014aza, Roberts:2015dea, Segovia:2015hra, Segovia:2016zyc, Lu:2017cln, Chen:2017pse}, on the baryon spectrum, and nucleon and $\Delta$ elastic and transition form factors, has been built upon this knowledge of the impact of omitting resonant contributions and the magnitude of their effects.  Therefore, a comparison between the empirical value of the Roper resonance pole-position and the computed dressed-quark core mass of the nucleon's radial excitation is not the critical test.  Instead, it is that between the masses of the quark core and the value determined for the meson-undressed bare-Roper, \emph{viz}.\ (in GeV):
\begin{equation}
\label{eqMassesA}
\begin{array}{l|ccc|c}
    & \mbox{R}_{{\rm core}}^{\mbox{\footnotesize \cite{Segovia:2015hra, Chen:2017pse}}}
    & \mbox{R}_{{\rm core}}^{\mbox{\footnotesize \cite{Lu:2017cln}}}
    & \mbox{R}_{{\rm core}}^{\mbox{\footnotesize \cite{Wilson:2011aa}}}
    & \mbox{R}_{\rm DCC\,bare}^{\mbox{\footnotesize \cite{Suzuki:2009nj}}} \\\hline
 \mbox{mass} & 1.73 & 1.82 & 1.72 & 1.76
\end{array}\,.
\end{equation}
Evidently, the DCC bare-Roper mass agrees with the quark core results obtained using both a QCD-kindred interaction \cite{Segovia:2015hra, Chen:2017pse} and refined treatments of a strictly-implemented vector$\,\otimes\,$vector contact-interaction \cite{Wilson:2011aa, Lu:2017cln}.\footnote{It is also commensurate with the value obtained in simulations of lattice-regularised QCD whose formulation and/or parameters suppress MB\,FSIs \cite{Mahbub:2010rm, Edwards:2011jj, Engel:2013ig, Liu:2014jua, Alexandrou:2014mka, Liu:2016rwa}.}
This is notable because all these calculations are independent, with just one common feature; namely, an appreciation that observed hadrons are built from a dressed-quark core plus a meson-cloud.


\section{Electromagnetic Currents}
\label{secEMcurrent}
With Faddeev amplitudes for the participating states in hand, computation of the desired elastic and transition form factors is a straightforward numerical exercise once the electromagnetic current is specified.  When the initial and final states are $I=1/2$, $J=1/2^+$ baryons, that current is completely specified by two form factors, \emph{viz}.
\begin{equation}
\bar u_{f}(P_f)\big[ \gamma_\mu^T F_{1}^{fi}(Q^2)+\frac{1}{m_{{fi}}} \sigma_{\mu\nu} Q_\nu F_{2}^{fi}(Q^2)\big] u_{i}(P_i)\,,
\label{NRcurrents}
\end{equation}
where: $u_{i}$, $\bar u_{f}$ are, respectively, Dirac spinors describing the incoming/outgoing baryons, with four-momenta $P_{i,f}$ and masses $m_{i,f}$ so that $P_{i,f}^2=-m_{i,f}^2$; $Q=P_f-P_i$; $m_{{fi}} = (m_f+m_{i})$; and $\gamma^T \cdot Q= 0$.

\begin{figure}[!t]
\centerline{\includegraphics[clip,width=1.0\linewidth]{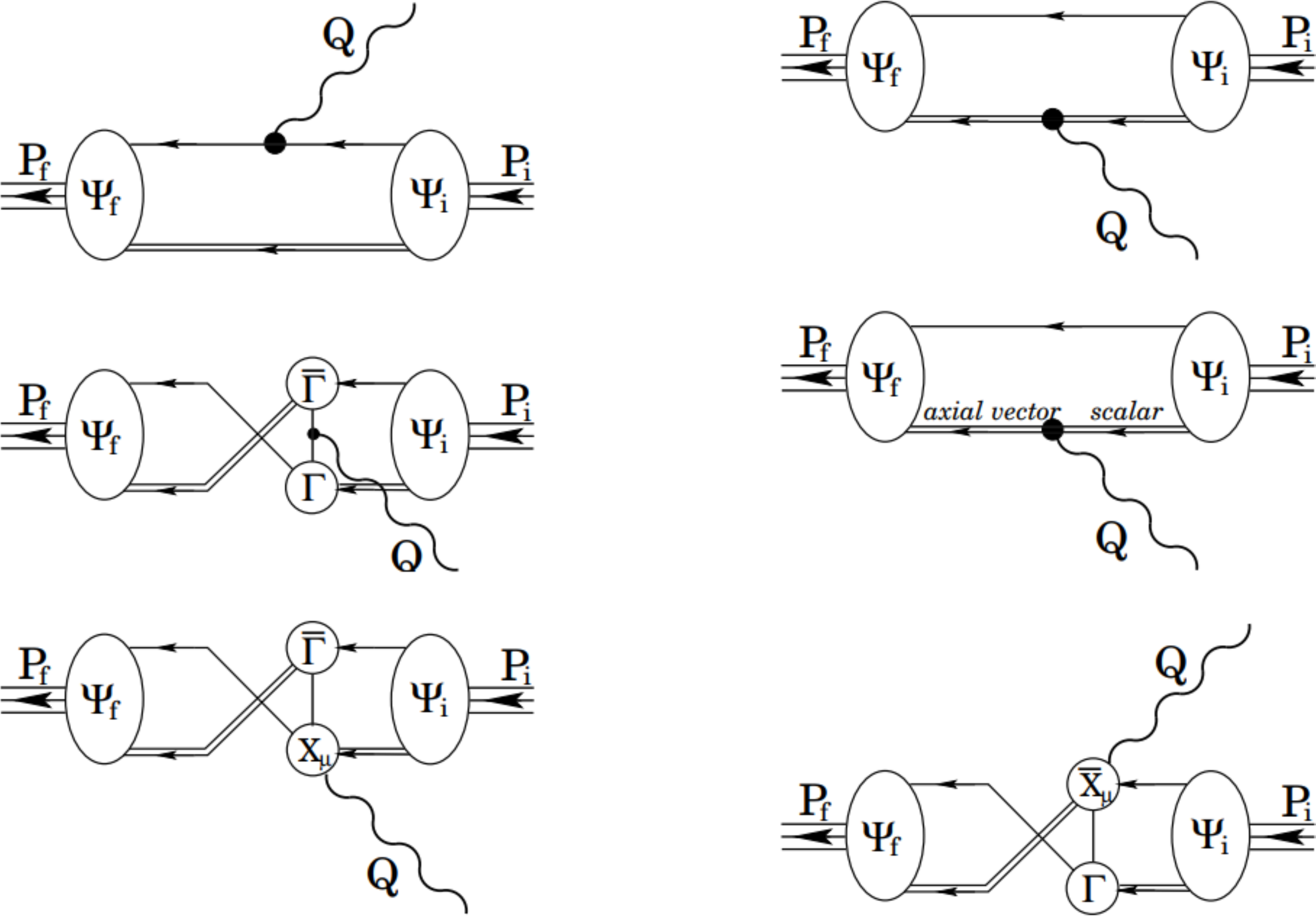}}
\caption{\label{vertexB}
Vertex that ensures a conserved current for on-shell baryons that are described by the Faddeev amplitudes produced by the equation depicted in Fig.\,\ref{figFaddeev}: \emph{single line}, dressed-quark propagator; \emph{undulating line}, photon; $\Gamma$,  diquark correlation amplitude; and \emph{double line}, diquark propagator.  Diagram~1 is the top-left image; the top-right is Diagram~2;
and so on, with Diagram~6 being the bottom-right image.
(Details are provided in Ref.\,\cite{Segovia:2014aza},  Appendix~C.)
}
\end{figure}

\begin{figure*}[!t]
\begin{center}
\begin{tabular}{lr}
\includegraphics[clip,width=0.43\linewidth]{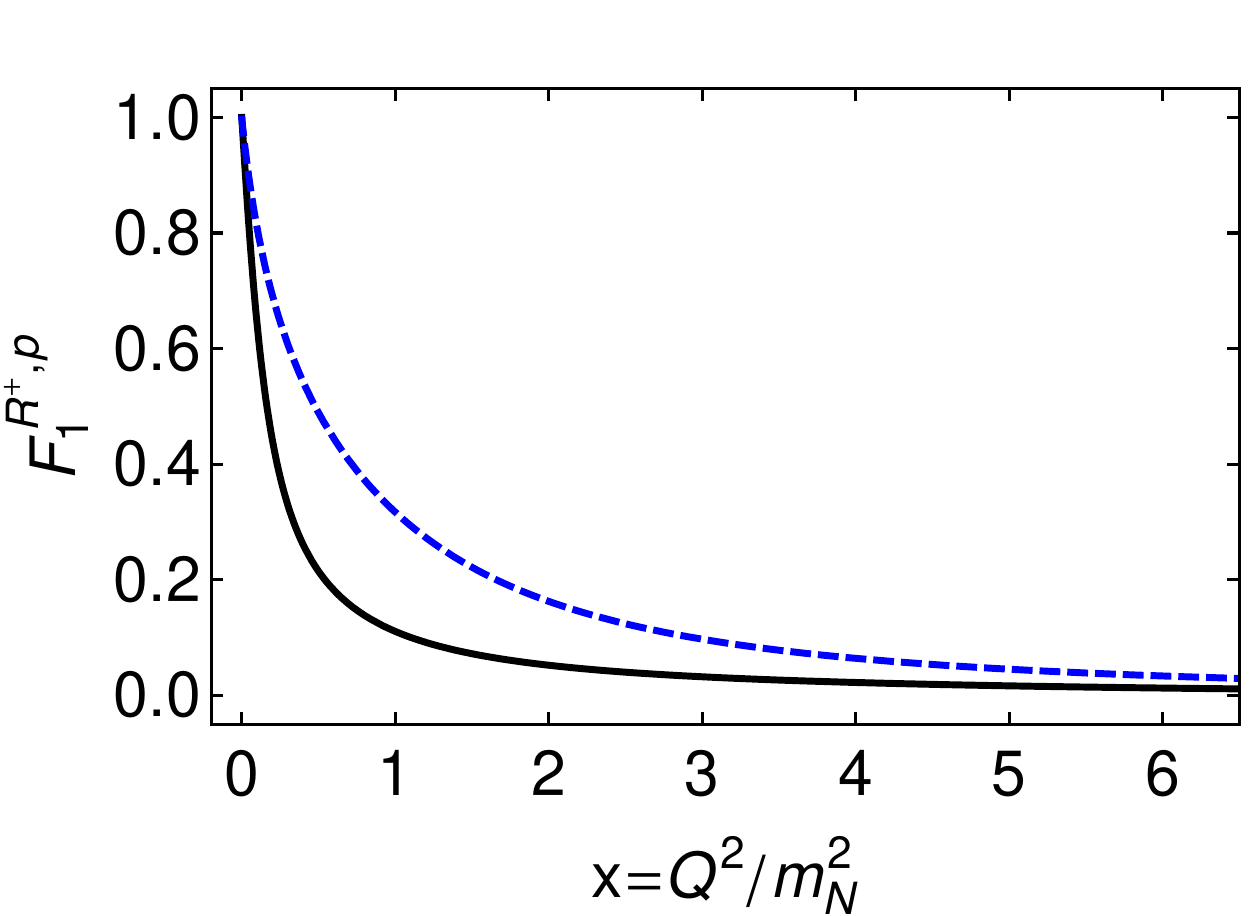}\hspace*{2ex } &
\includegraphics[clip,width=0.43\linewidth, height=0.24\textheight]{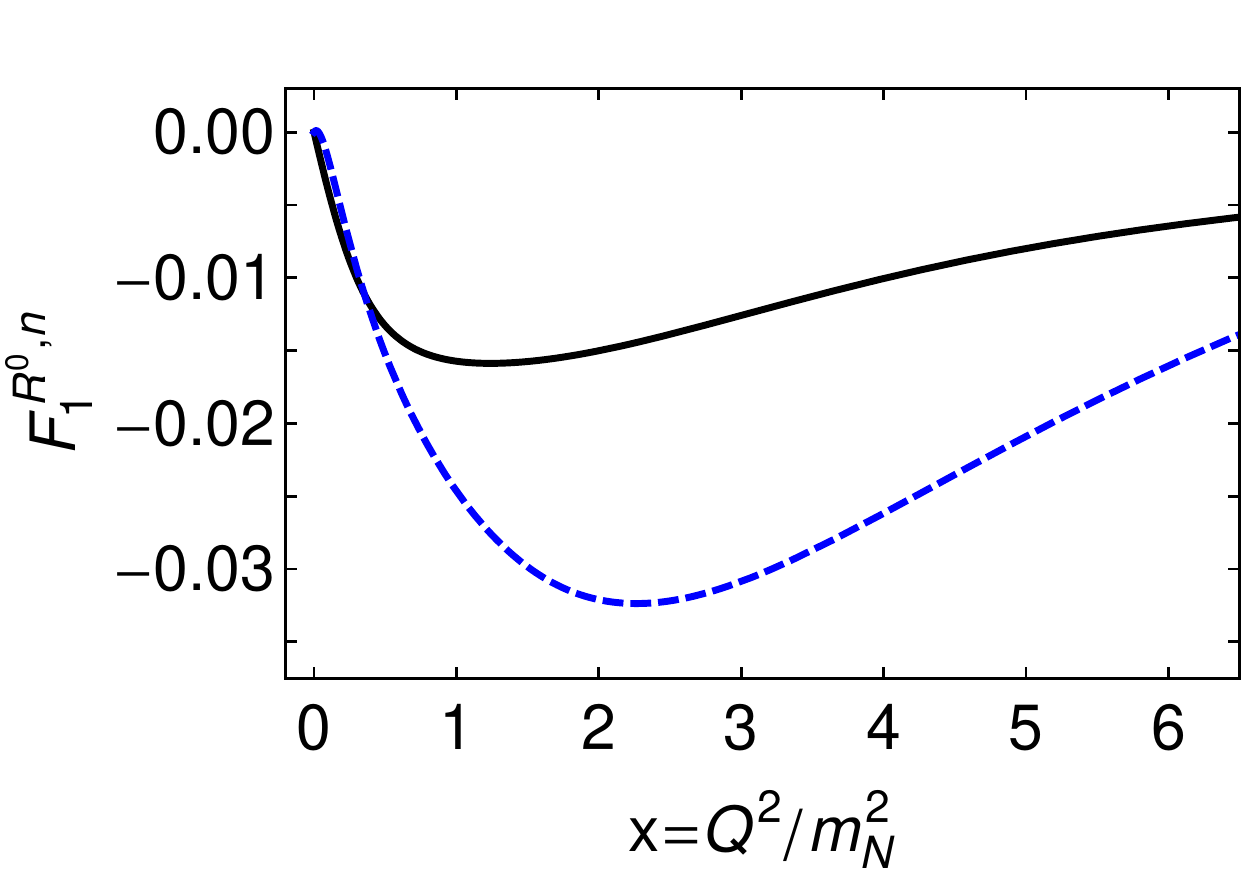}\vspace*{-0ex}
\end{tabular}
\begin{tabular}{lr}
\includegraphics[clip,width=0.43\linewidth]{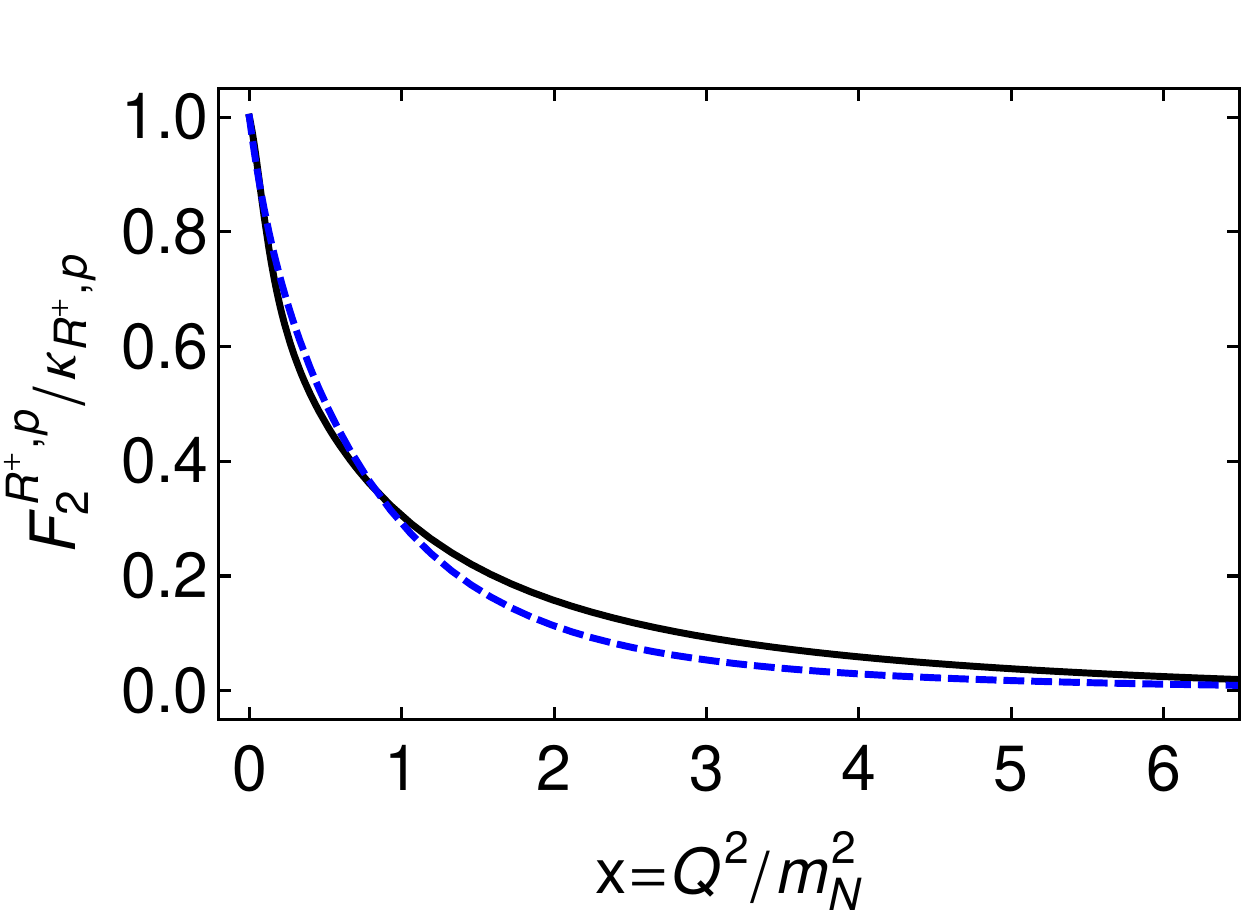}\hspace*{2ex } &
\includegraphics[clip,width=0.43\linewidth, height=0.24\textheight]{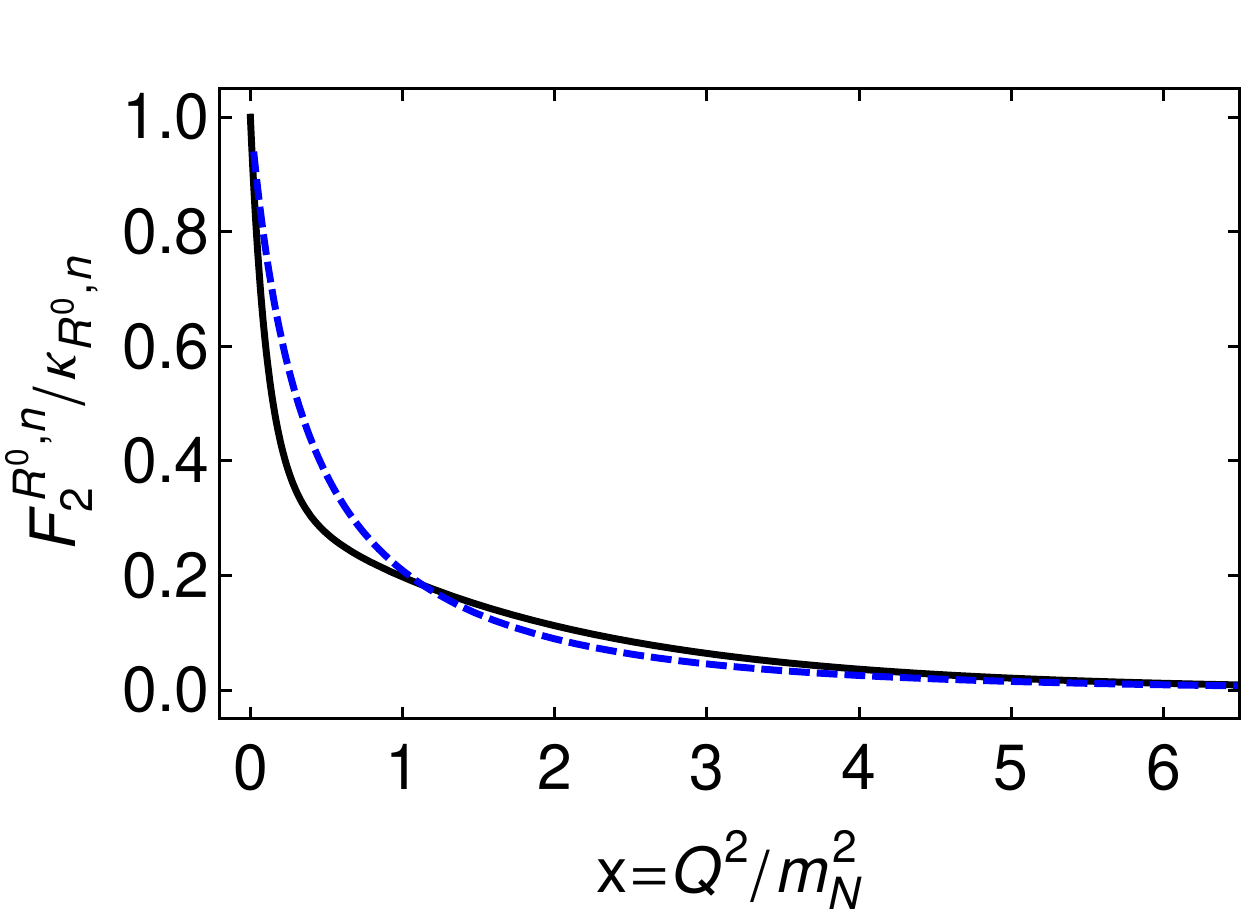}\vspace*{-1ex}
\end{tabular}
\end{center}
\caption{\label{elastic}
Solid (black) curves -- Dirac (upper panels) and Pauli (lower) elastic electromagnetic form factors associated with the dressed-quark cores of the charged (left) and neutral (right) Roper systems.
Dashed (blue) curves -- analogous results for proton and neutron.
($\kappa_{R,N} = F_2^{R,N}(x=0)$; $x=Q^2/m_N^2$, where $M_N = 1.18\,$GeV is the nucleon' dressed-quark core mass.)
}
\end{figure*}

The vertex sufficient to express the interaction of a photon with a baryon generated by the Faddeev equation in Fig.\,\ref{figFaddeev} is described elsewhere \cite{Oettel:1999gc, Segovia:2014aza}.  It is a sum of six terms, depicted in Fig.\,\ref{vertexB}, with the photon separately probing the quarks and diquarks in various ways, so that diverse features of quark dressing and the quark-quark correlations all play a role in determining the form factors.  To elaborate, elastic and transition electromagnetic form factors involving the nucleon and Roper may be dissected in two separate ways, each of which can be considered as a sum of three distinct terms, \emph{viz}.
\begin{description}
\item[DD = diquark dissection]$\,$\\[-3.5ex]
\begin{description}
\setlength\itemsep{0em}
\item[\mbox{\emph{DD1}}] sca\-lar diquark, $[ud]$, in both the initial- and final-state baryon,
%
\item[\emph{DD2}] pseudovector diquark, $\{qq\}$, in both the initial- and final-state baryon, and
\item[\emph{DD3}] a different diquark in the initial- and final-state baryon;
\end{description}
\item[DS = scatterer dissection]$\,$\\[-3.5ex]
\begin{description}
\setlength\itemsep{0em}
\item[\emph{DS1}] photon strikes a bystander dressed-quark (Diagram~1 in Fig.\,\ref{vertexB}),
\item[\emph{DS2}] photon interacts with a diquark, elastically or causing a transition scalar\,$\leftrightarrow$\,pseudovector (Diagrams~2 and 4 in Fig.\,\ref{vertexB}), and
\item[\emph{DS3}] photon strikes a dressed-quark in-flight, as one diquark breaks up and another is formed (Diagram ~3 in Fig.\,\ref{vertexB}), or appears in one of the two associated ``seagull'' terms (Diagrams~5 and 6).
\end{description}
\end{description}
The anatomy of a given transition is revealed by merging the information provided by DD and DS.

\begin{figure*}[!t]
\begin{center}
\begin{tabular}{lr}
\includegraphics[clip,width=0.43\linewidth]{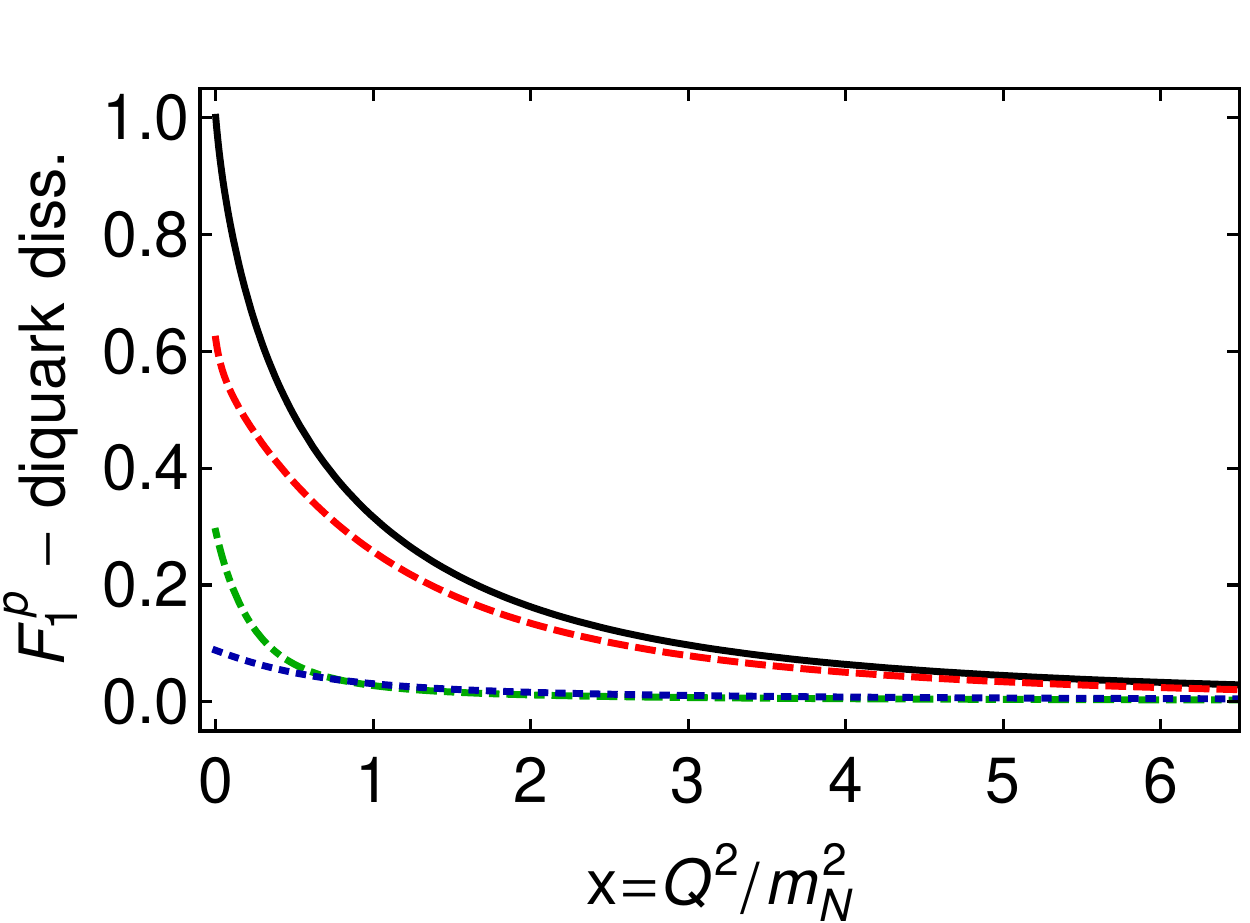}\hspace*{2ex } &
\includegraphics[clip,width=0.43\linewidth, height=0.24\textheight]{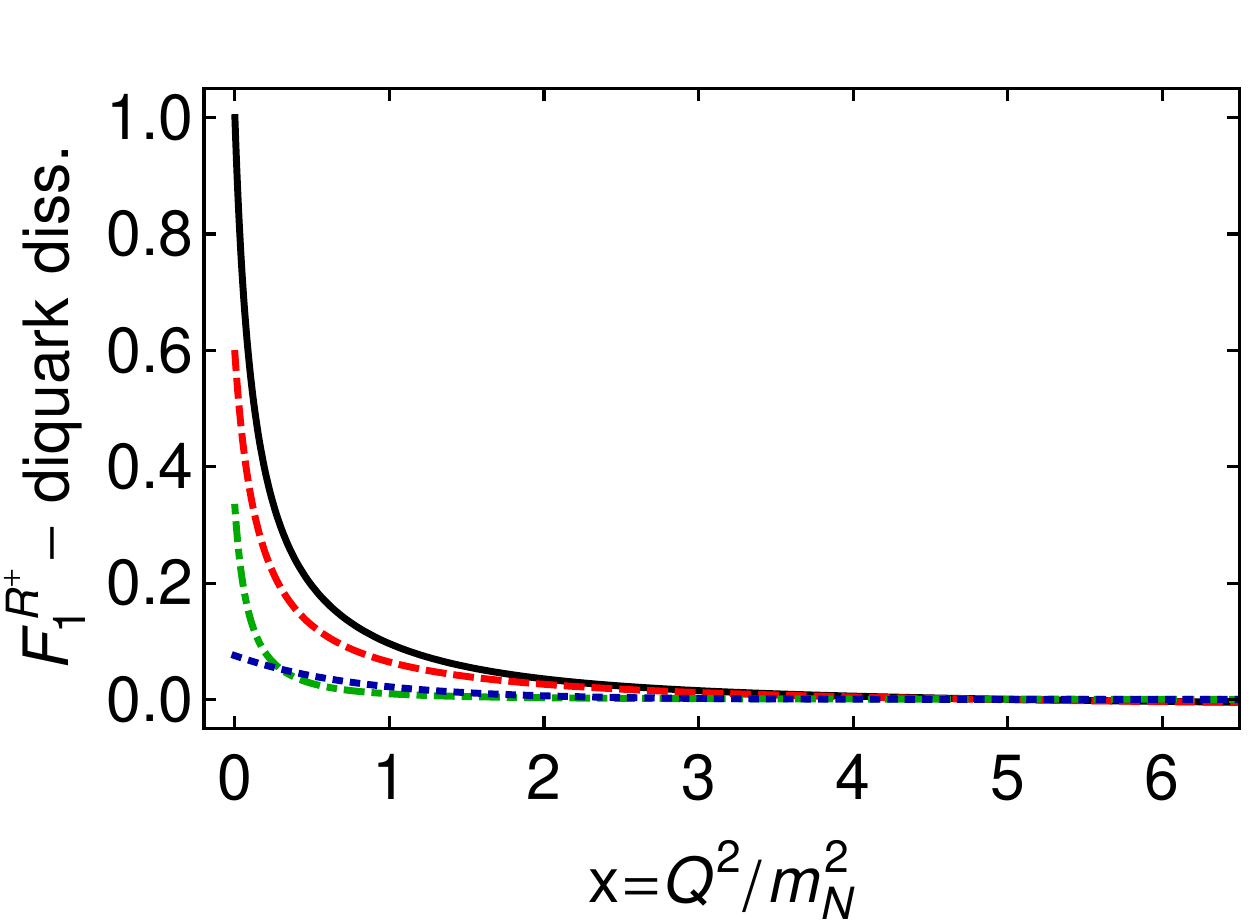}\vspace*{-0ex}
\end{tabular}
\begin{tabular}{lr}
\includegraphics[clip,width=0.43\linewidth]{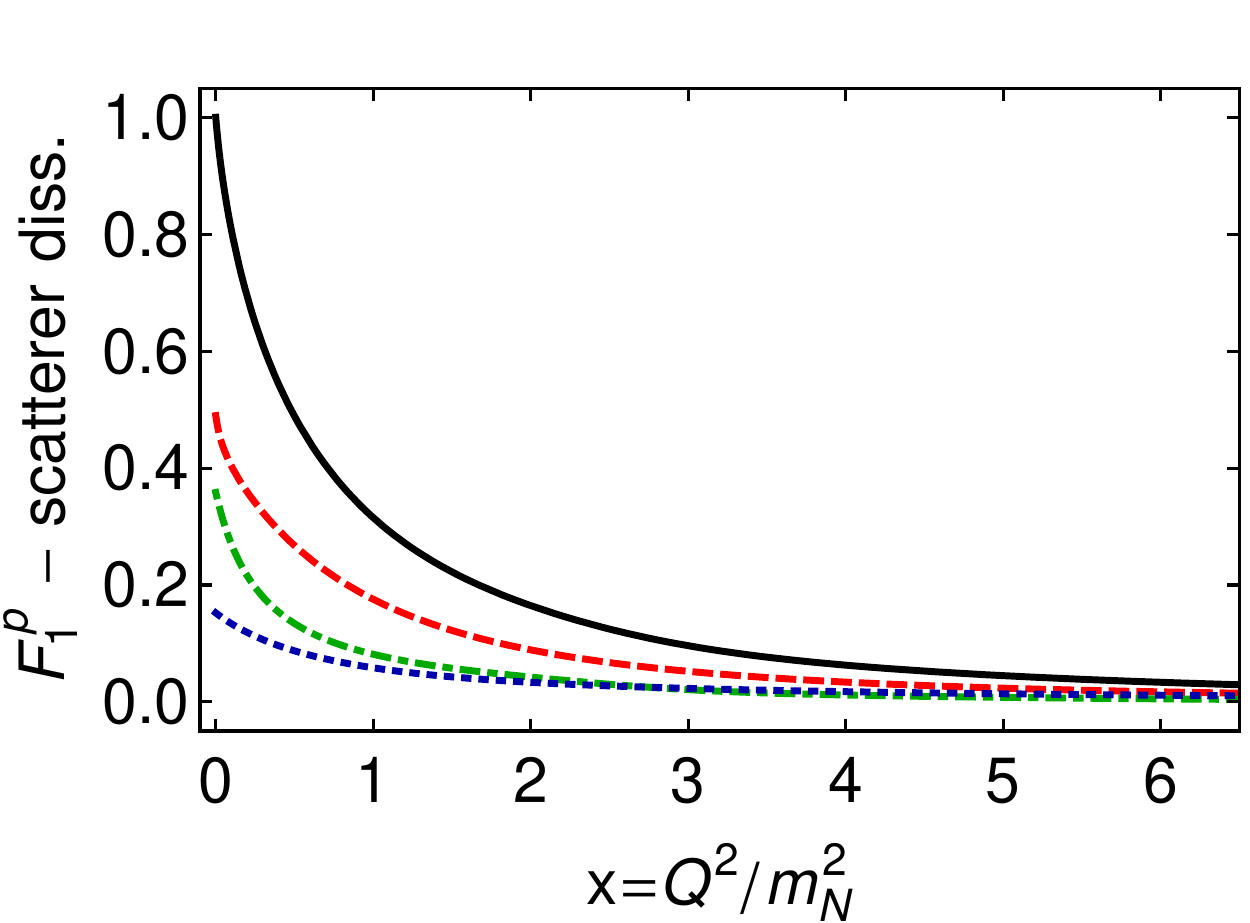}\hspace*{2ex } &
\includegraphics[clip,width=0.43\linewidth, height=0.24\textheight]{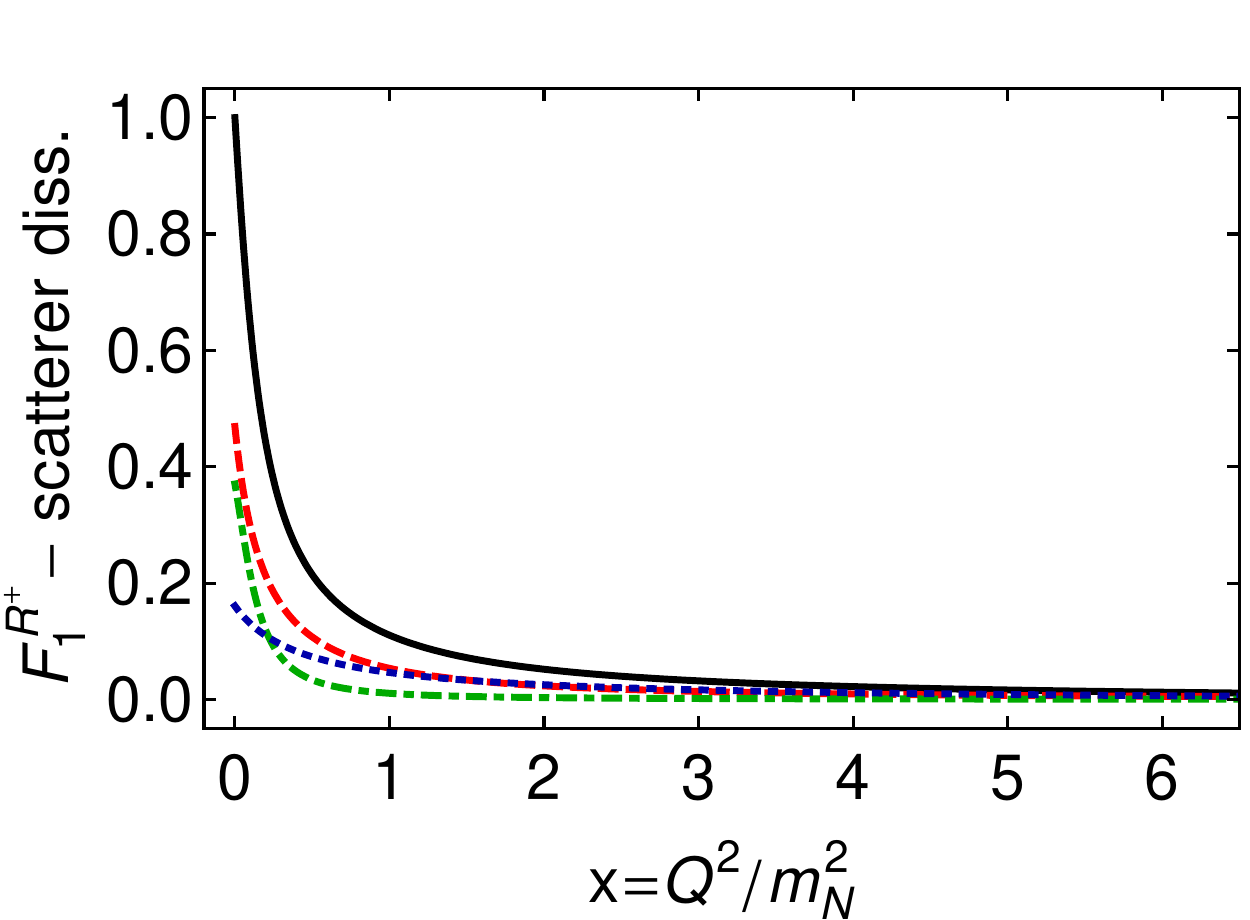}\vspace*{-1ex}
\end{tabular}
\end{center}
\caption{\label{elasticdissectF1}
Dirac form factors of the proton (left) and charged-Roper (right).
\emph{Upper panels} -- diquark breakdown: \emph{DD1} (dashed red), scalar diquark in initial and final baryon; \emph{DD2} (dot-dashed green), pseudovector diquark in both initial and final states; \emph{DD3} (dotted blue), scalar diquark in incoming baryon, pseudovector diquark in outgoing baryon, and vice versa.
\emph{Lower panels} -- scatterer breakdown: \emph{DS1} (red dashed), photon strikes an uncorrelated dressed quark; \emph{DS2} (dot-dashed green), photon strikes a diquark; and \emph{DS3} (dotted blue), diquark breakup contributions, including photon striking exchanged dressed-quark.
}
\end{figure*}

\begin{figure*}[!t]
\begin{center}
\begin{tabular}{lr}
\includegraphics[clip,width=0.43\linewidth]{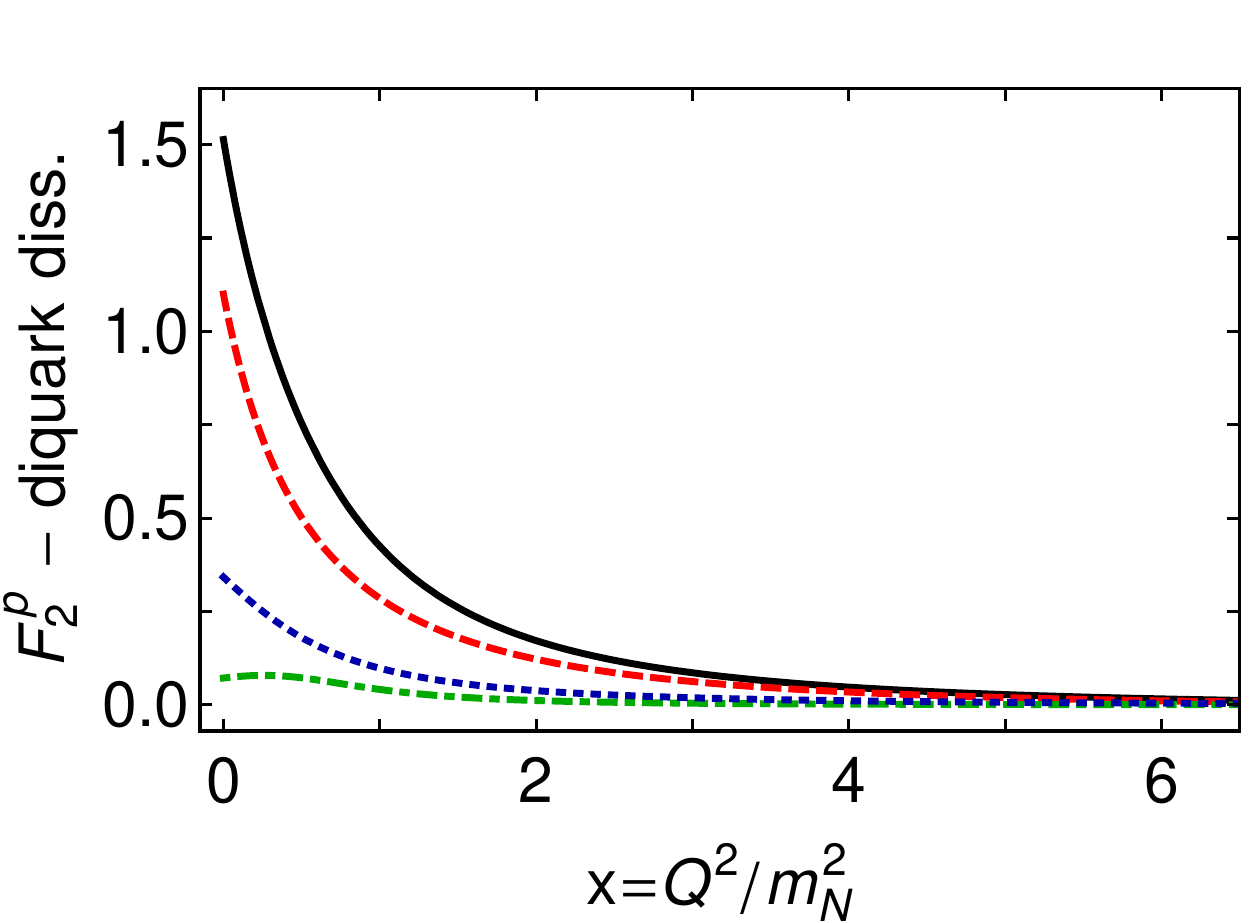}\hspace*{2ex } &
\includegraphics[clip,width=0.43\linewidth, height=0.24\textheight]{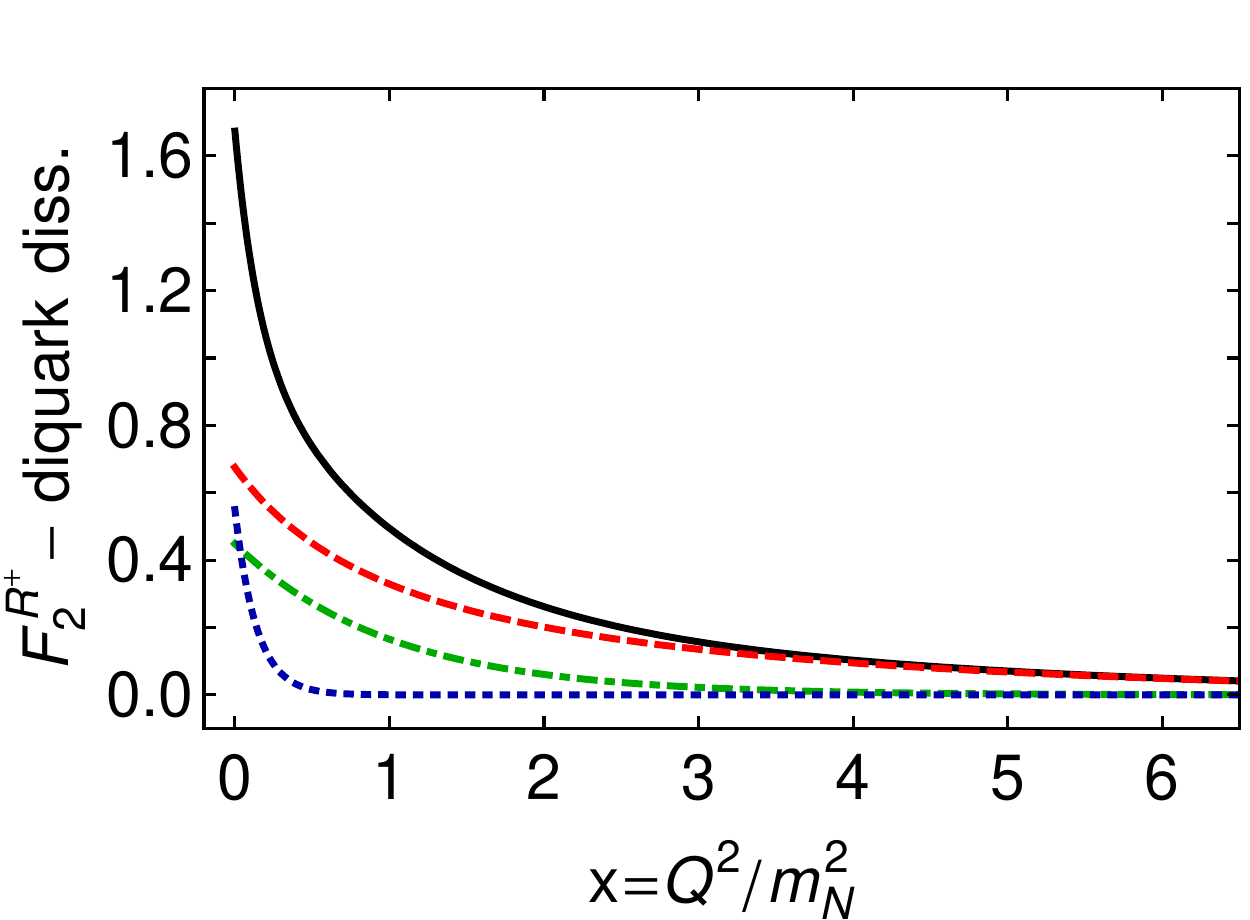}\vspace*{-0ex}
\end{tabular}
\begin{tabular}{lr}
\includegraphics[clip,width=0.43\linewidth]{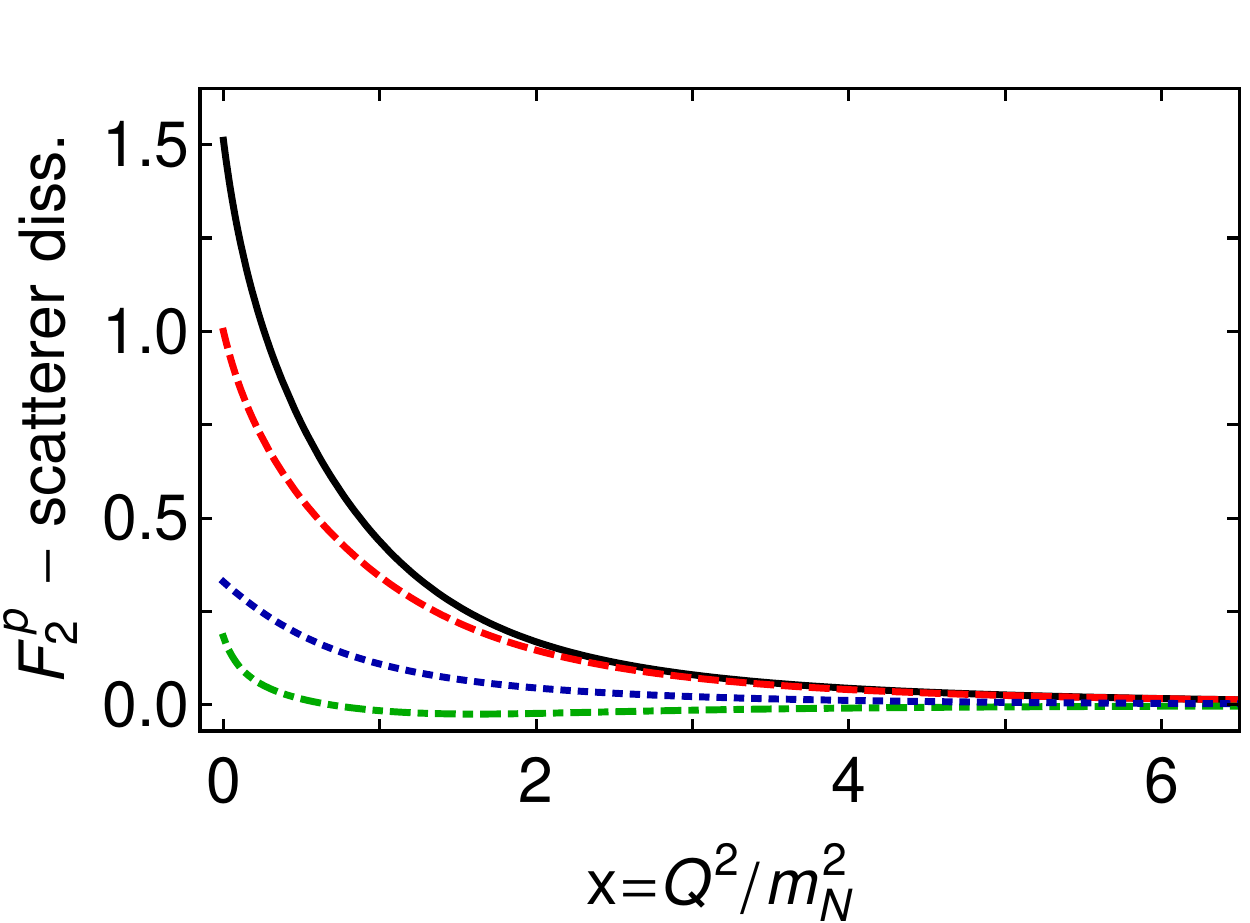}\hspace*{2ex } &
\includegraphics[clip,width=0.43\linewidth, height=0.24\textheight]{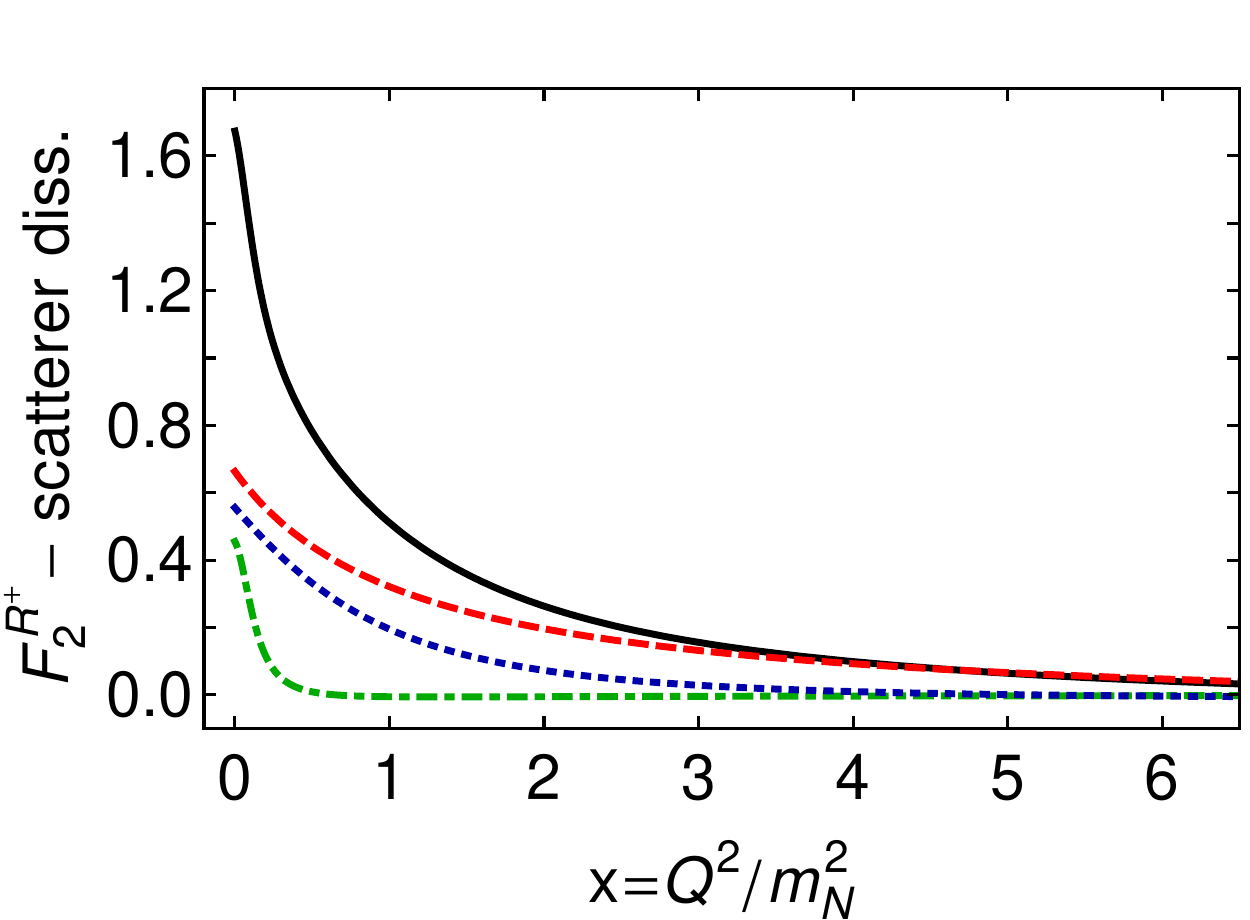}\vspace*{-1ex}
\end{tabular}
\end{center}
\caption{\label{elasticdissectF2}
Pauli form factors of the proton and charged-Roper.
\emph{Upper panels} -- diquark breakdown: \emph{DD1} (dashed red), scalar diquark in initial and final baryon; \emph{DD2} (dot-dashed green), pseudovector diquark in both initial and final states; \emph{DD3} (dotted blue), scalar diquark in incoming baryon, pseudovector diquark in outgoing baryon, and vice versa.
\emph{Lower panels} -- scatterer breakdown: \emph{DS1} (red dashed), photon strikes an uncorrelated dressed quark; \emph{DS2} (dot-dashed green), photon strikes a diquark; and \emph{DS3} (dotted blue), diquark breakup contributions, including photon striking exchanged dressed-quark.
}
\end{figure*}

\begin{figure*}[!t]
\begin{center}
\begin{tabular}{lr}
\includegraphics[clip,width=0.43\linewidth]{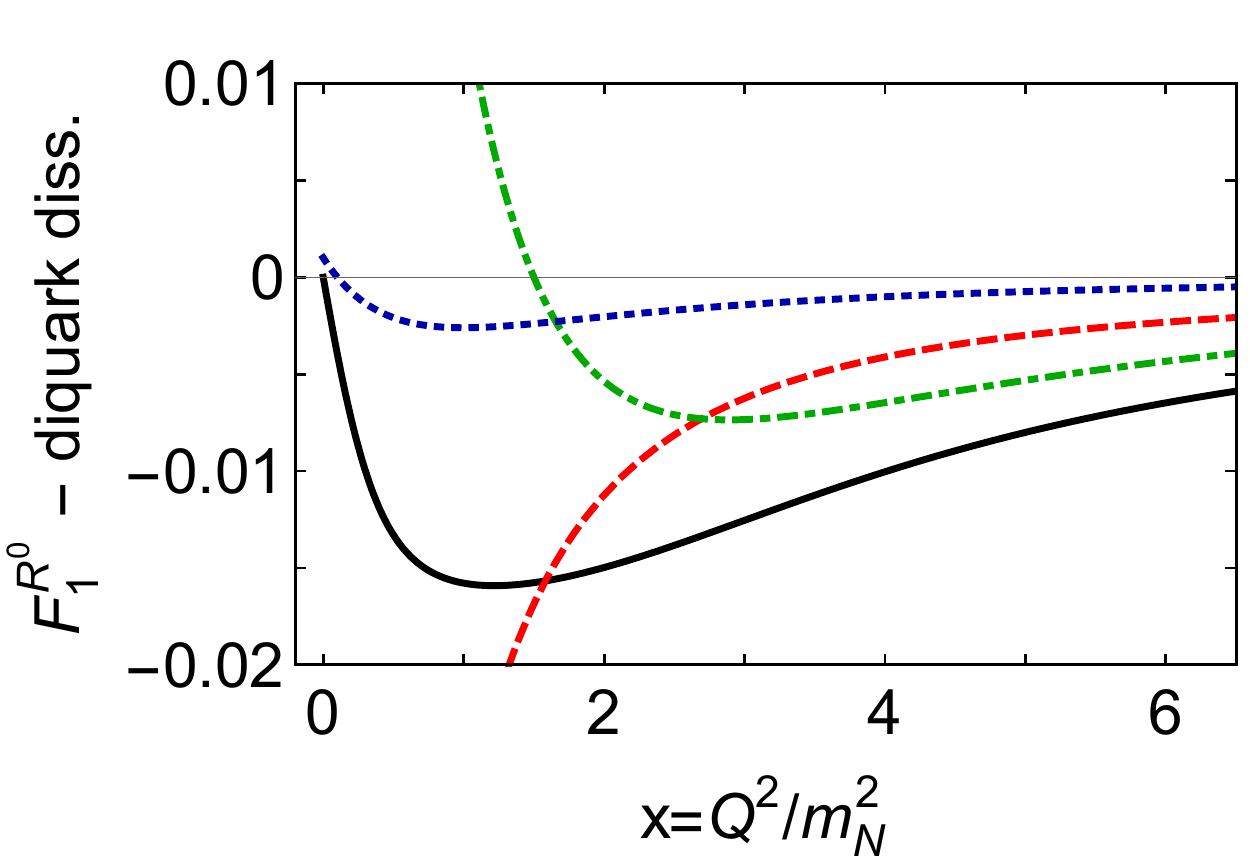}\hspace*{2ex } &
\includegraphics[clip,width=0.43\linewidth, height=0.225\textheight]{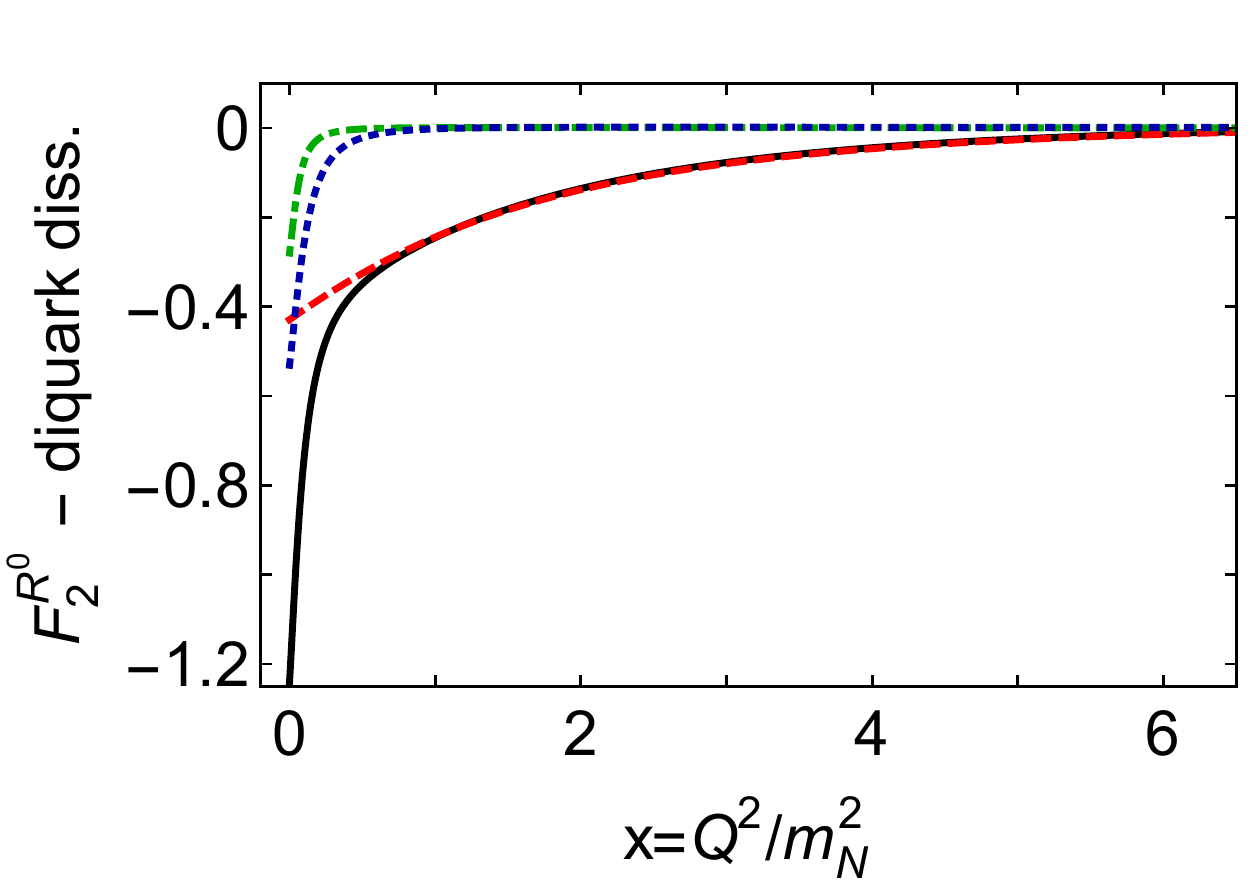}\vspace*{-0ex}
\end{tabular}
\begin{tabular}{lr}
\includegraphics[clip,width=0.43\linewidth]{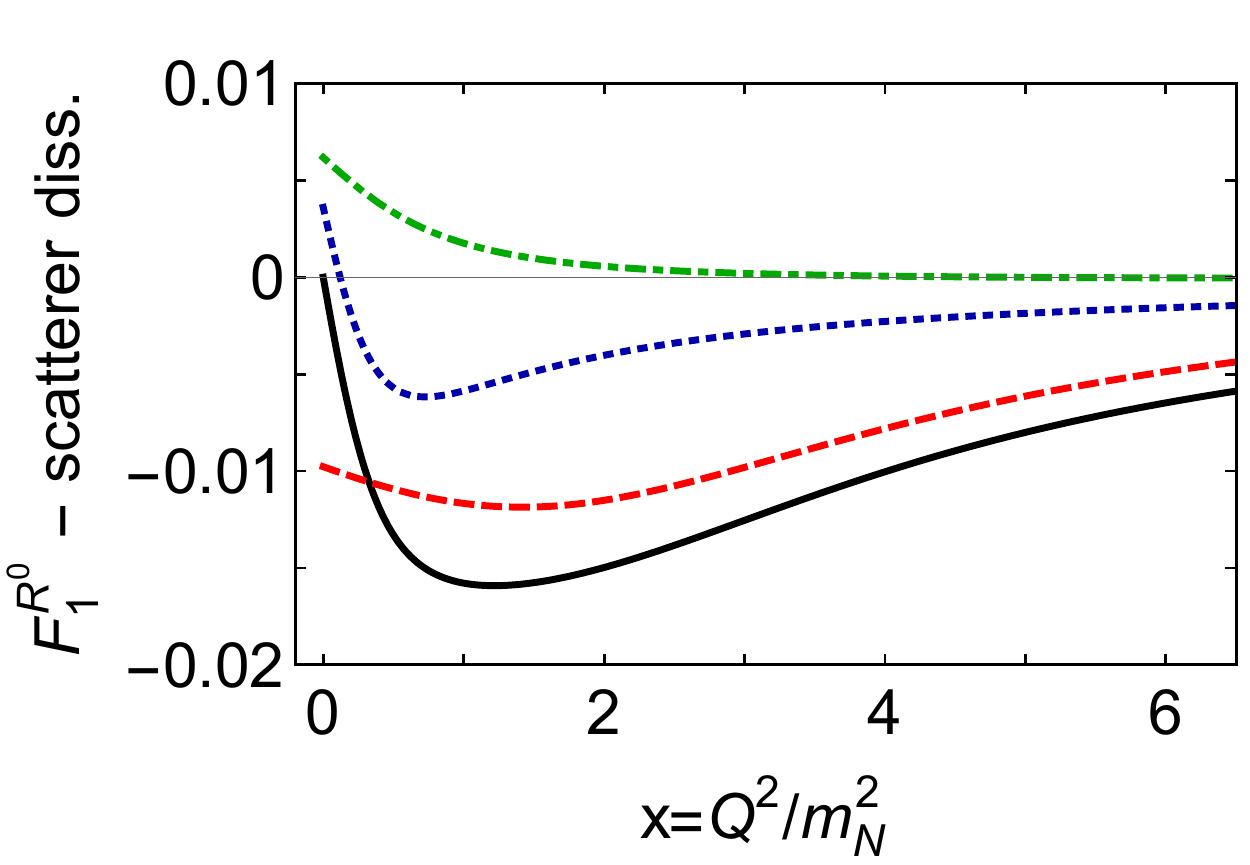}\hspace*{2ex } &
\includegraphics[clip,width=0.43\linewidth, height=0.225\textheight]{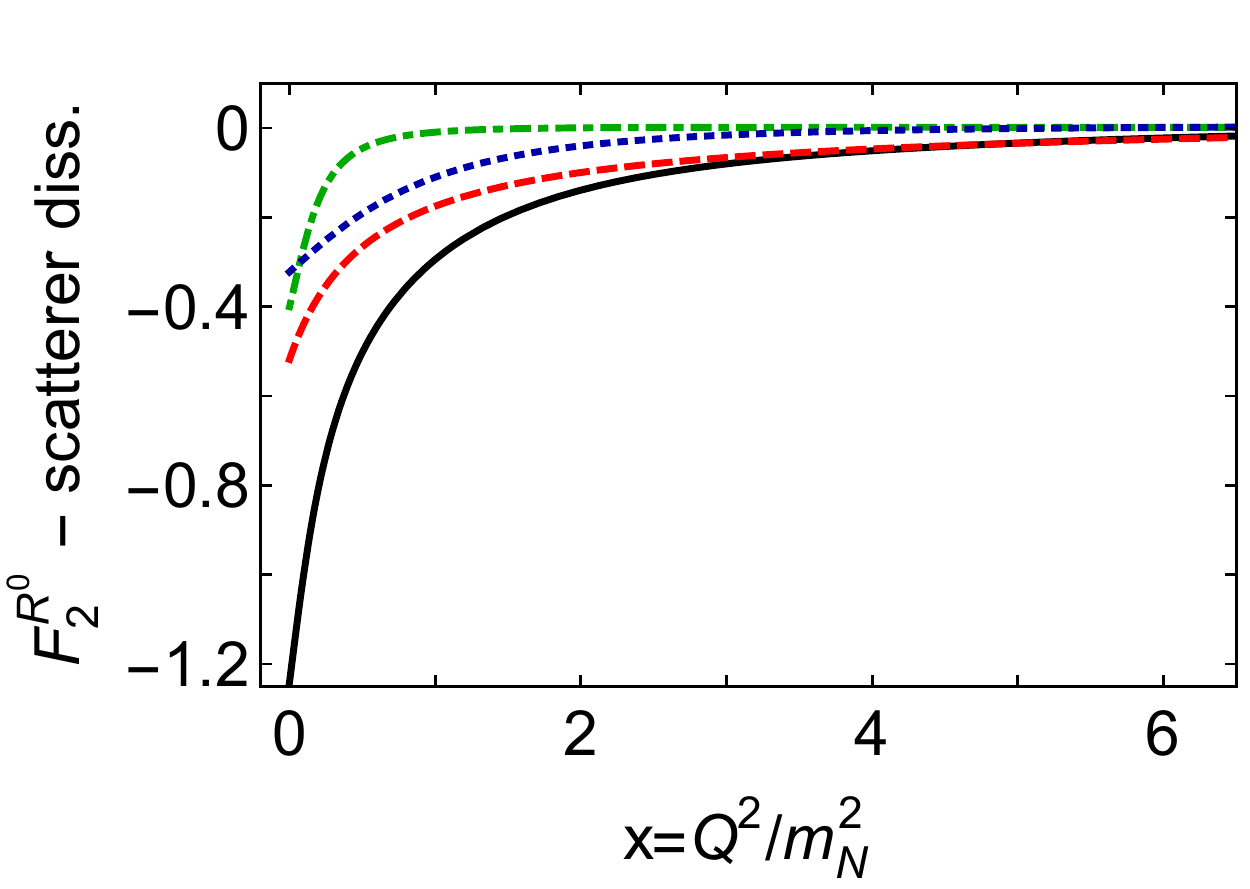}\vspace*{-1ex}
\end{tabular}
\end{center}
\caption{\label{elasticdissectF12neutral}
Dirac (left) and Pauli (right) elastic form factors of neutral-Roper.
\emph{Upper panels} -- diquark breakdown: \emph{DD1} (dashed red), scalar diquark in initial and final baryon; \emph{DD2} (dot-dashed green), pseudovector diquark in both initial and final states; \emph{DD3} (dotted blue), scalar diquark in incoming baryon, pseudovector diquark in outgoing baryon, and vice versa.
\emph{Lower panels} -- scatterer breakdown: \emph{DS1} (red dashed), photon strikes an uncorrelated dressed quark; \emph{DS2} (dot-dashed green), photon strikes a diquark; and \emph{DS3} (dotted blue), diquark breakup contributions, including photon striking exchanged dressed-quark.
}
\end{figure*}

\section{Elastic Form Factors}
\label{secEMelastic}
In any computation of transition form factors, one must first calculate the analogous elastic form factors for the states involved because the associated values of $F_1^{f=i}(Q^2=0)$ fix the normalisation of the transition.  The elastic Dirac and Pauli form factors associated with the dressed-quark core of the charged and neutral Roper are depicted in Fig.\,\ref{elastic} and compared with those for the proton and neutron from Ref.\,\cite{Segovia:2014aza}.  Evidently, there are qualitative similarities and quantitative differences.  The latter are seemingly prominent in the $Q^2$-dependence of the Dirac form factors of the neutral-Roper and neutron; but here appearances are deceptive because both functions are independently computed as the valence-quark electric-charge-weighted sum of larger, positive quantities, with cancellations leading to uniformly small results.

\begin{table}[b]
\caption{\label{tabstatic}
Static properties derived from the elastic form factors depicted in Fig.\,\ref{elastic}, see Eq.\,\eqref{eqradii} and following text.
($M_N = 1.18\,$GeV is the nucleon dressed-quark core mass.)}
\begin{center}
\begin{tabular*}
{\hsize}
{
l|@{\extracolsep{0ptplus1fil}}
c@{\extracolsep{0ptplus1fil}}
c@{\extracolsep{0ptplus1fil}}
c@{\extracolsep{0ptplus1fil}}
c@{\extracolsep{0ptplus1fil}}}\hline
  & $R^+$ & $p$ & $R^0$ & $n$ \\\hline
$r_E \, M_N$ & 6.23 & 3.65 & $\phantom{-}0.93i$ & $\phantom{-}1.67i$\\
$r_M \, M_N$ & 4.49 & 3.17 & $\phantom{-}4.15\phantom{i}$ & $\phantom{-}4.19\phantom{i}$\\
$\mu$ & 2.67 & 2.50 & $-1.24\phantom{i}$ & $-1.83\phantom{i}$ \\\hline
\end{tabular*}
\end{center}
\end{table}

Defining (Sachs) electric and magnetic form factors:
\begin{align}
G_E = F_1 - \frac{Q^2}{4 m_B^2} F_2\,,\quad
G_M = F_1 + F_2\,,
\end{align}
where $m_B$ is the baryon's mass, the $Q^2=0$ values and slopes of the form factors in Fig.\,\ref{elastic} yield the static properties listed in Table\,\ref{tabstatic}, where the radii are defined via
\begin{align}
\label{eqradii}
r^2 & = - \left.\frac{6}{\mathpzc n} \frac{d}{ dQ^2} G(Q^2)\right|_{Q^2=0}\,,
\end{align}
with ${\mathpzc n} = G(Q^2=0)$ when this quantity is nonzero, ${\mathpzc n} = 1$ otherwise, and the anomalous magnetic moment $\mu = G_M(0)$.
The electromagnetic radii of the charged-Roper core are larger than those of the proton core, but the magnetic moments are similar; and this pattern is reversed in the neutral-Roper/neutron comparison.

In order to reveal more details about the structural similarities and differences between the dressed-quark cores of the nucleon and Roper, in Figs.\,\ref{elasticdissectF1}, \ref{elasticdissectF2} we contrast the diquark and scatterer dissections of the Dirac and Pauli form factors, respectively, of the proton and $R^+$.  (The neutron/$R^0$ comparisons do not contain additional information because the associated form factors are simply different charge-weighted sums of the same basic contributions.)
Focusing first on Fig.\,\ref{elasticdissectF1}, it is apparent that every contribution to the $R^+$ elastic Dirac form factor falls more rapidly than its analogue in the proton.  On the other hand, the relative importance of each is typically the same within the proton and $R^+$, \emph{e.g}.\ in both cases, the DD1 term (scalar diquark, $[ud]$, in both initial and final state) dominates the Dirac form factor, with the largest contribution arising from the photon striking the bystander quark (DS1).  The former observation highlights that the relative strengths of the various diquarks in both the nucleon and Roper are almost identical, with
\begin{equation}
{\cal P}_{[ud]}^{R^+} = 0.60 \,, \quad {\cal P}_{\{qq\}}^{R^+} = 0.33 \,, \quad {\cal P}_{\rm mix}^{R^+} = 0.07 \,
\end{equation}
where these probabilities are the $Q^2=0$ values of the individual DD terms, for the reasons explained elsewhere \cite{Cloet:2008re}.  The values for the proton are \cite{Segovia:2014aza}: $0.62$, $0.29$, $0.09$.

Turning to Fig.\,\ref{elasticdissectF2}, the primary contribution to the Pauli form factor in both cases is again delivered by the photon striking a bystander quark in association with $[ud]$ (DD1$\times$DS1 is dominant in both columns).  On the other hand, there are differences between subleading terms, \emph{e.g}.\ DD3$\times$DS1 in the proton is of similar importance to DD2$\times$DS1 in the $R^+$.  Since diagrams with pseudovector diquark spectators do not contribute to the charged-particle form factors considered herein when isospin symmetry is assumed \cite{Wilson:2011aa}, then this comparison indicates that, in the Pauli elastic form factor, the strength of $[ud]$-$\{qq\}$ transitions in the proton, with a photon striking the exchanged quark, is roughly matched by analogous $\{qq\}$-$\{qq\}$ breakup-recombination effects in the $R^+$.

For completeness and because the results can provide counterpoints for the $\gamma n \to R^0$ form factors, in Fig.\,\ref{elasticdissectF12neutral} we depict the diquark and scatterer dissections for the Dirac and Pauli elastic form factors of the neutral-Roper.

\section{Nucleon-to-Roper Transition}
\label{secEMtransition}
\begin{figure*}[!t]
\begin{center}
\begin{tabular}{cc}
\includegraphics[clip,width=0.43\linewidth]{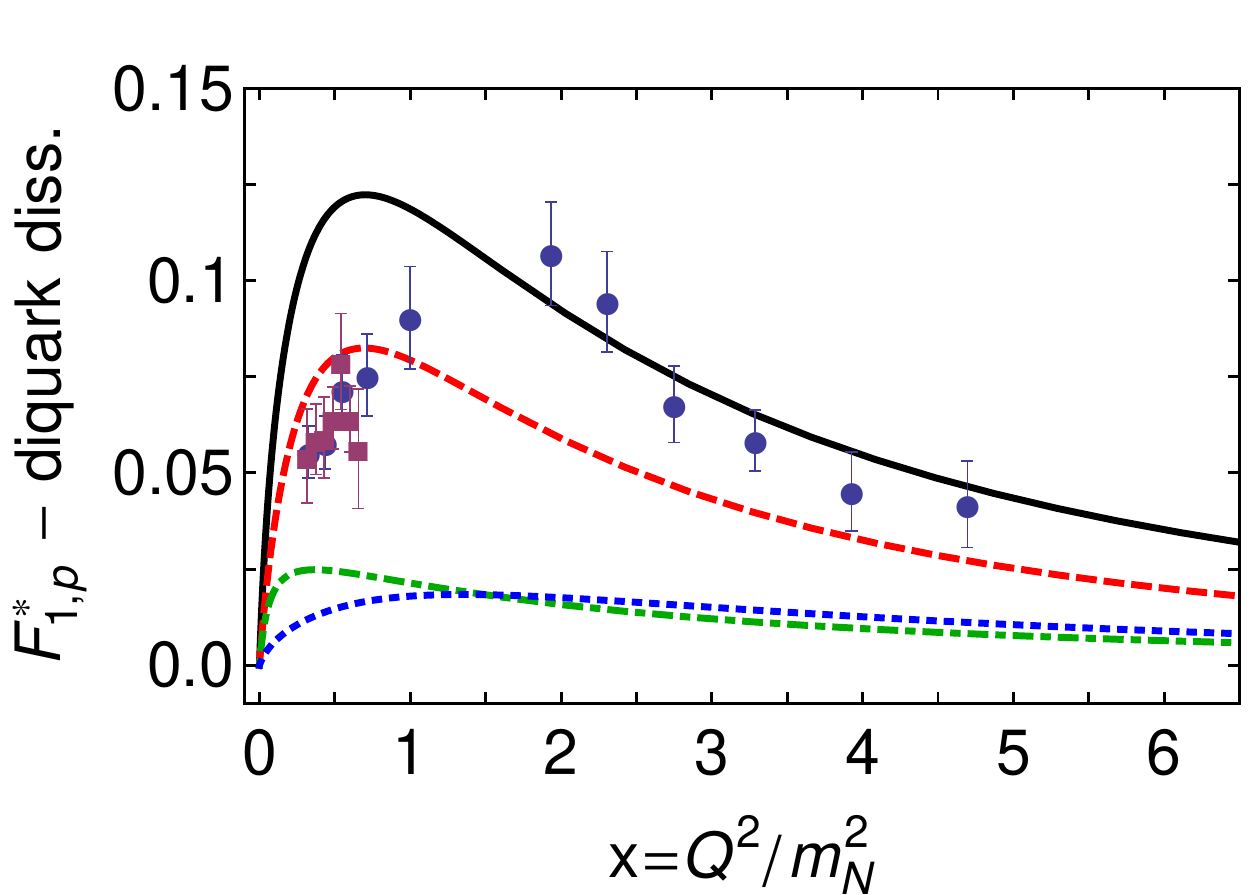}\hspace*{2ex } &
\includegraphics[clip,width=0.43\linewidth, height=0.235\textheight]{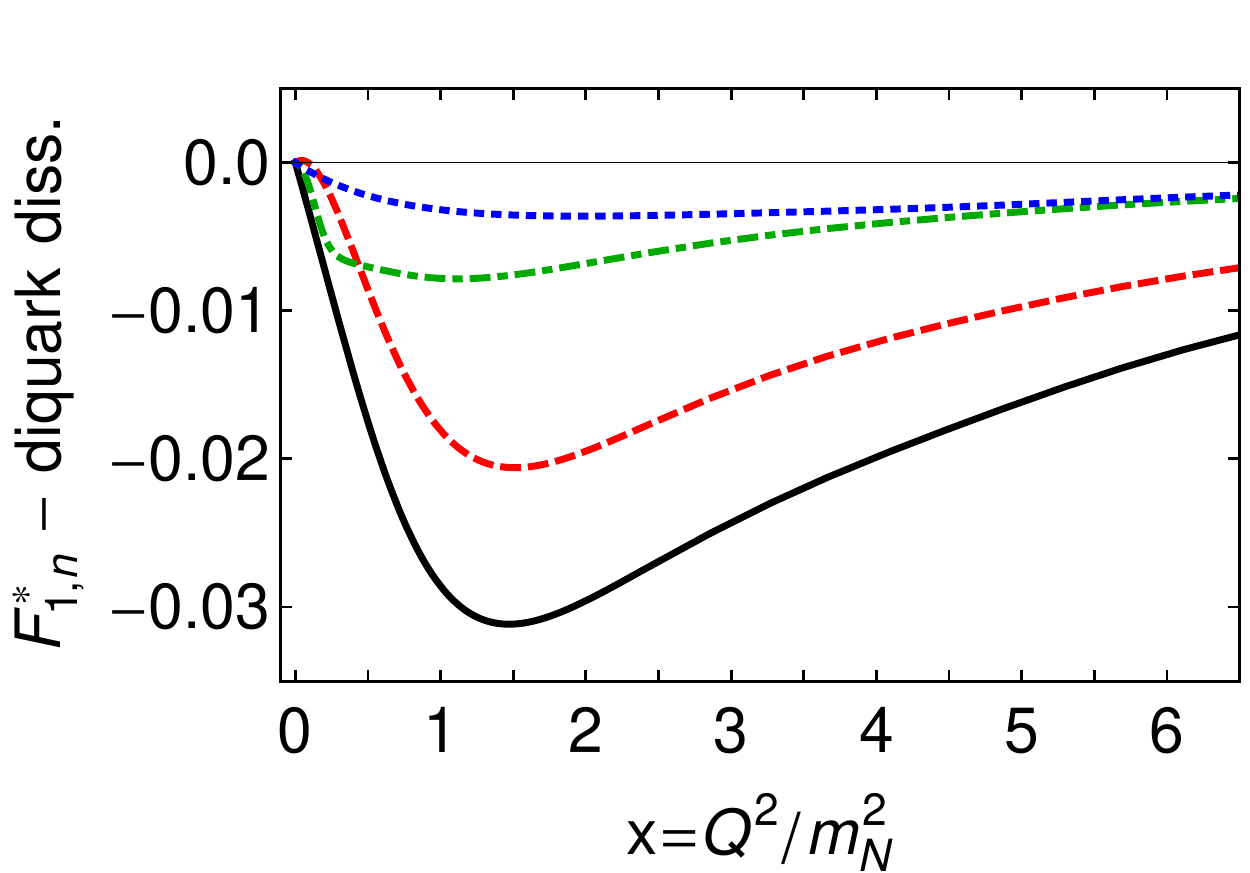}\vspace*{-0ex}
\end{tabular}
\begin{tabular}{cc}
\includegraphics[clip,width=0.43\linewidth]{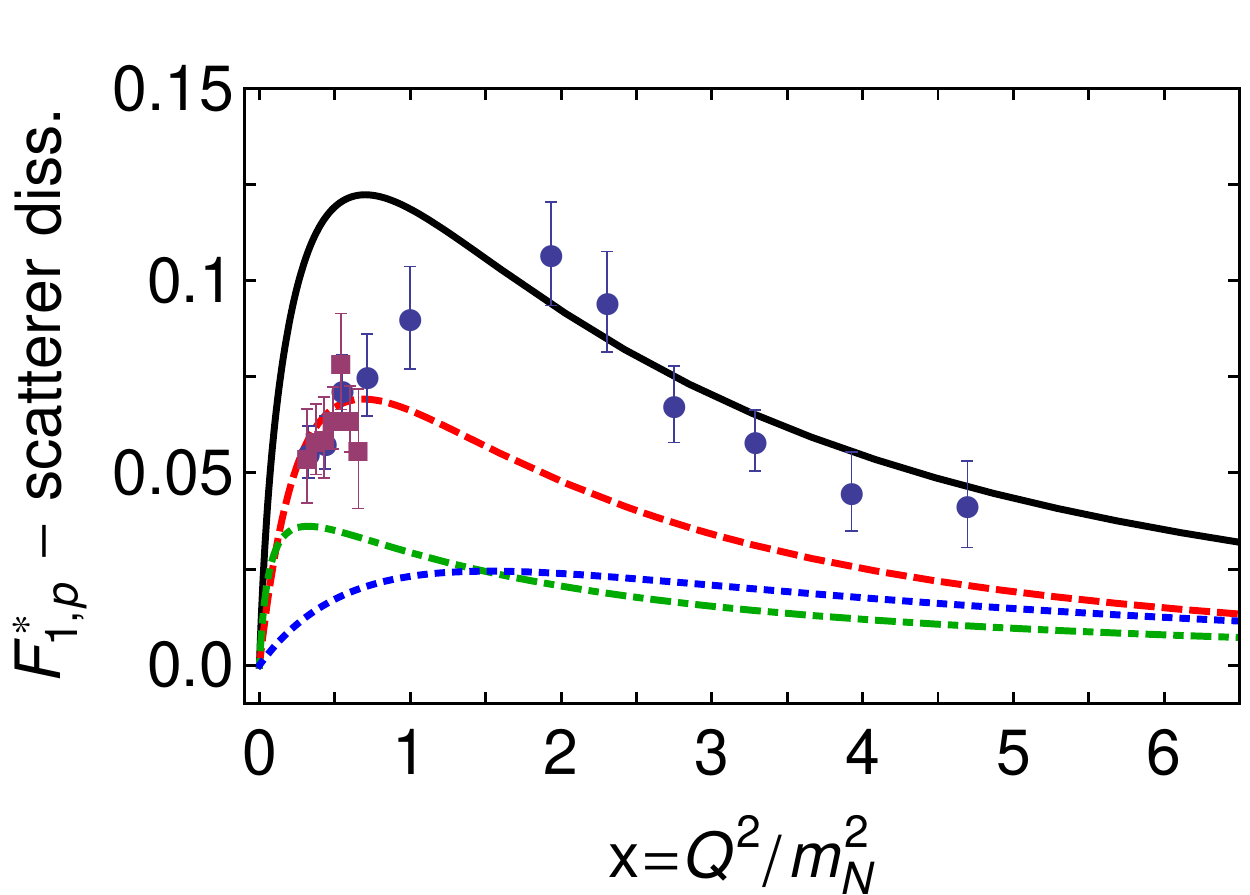}\hspace*{2ex } &
\includegraphics[clip,width=0.43\linewidth, height=0.235\textheight]{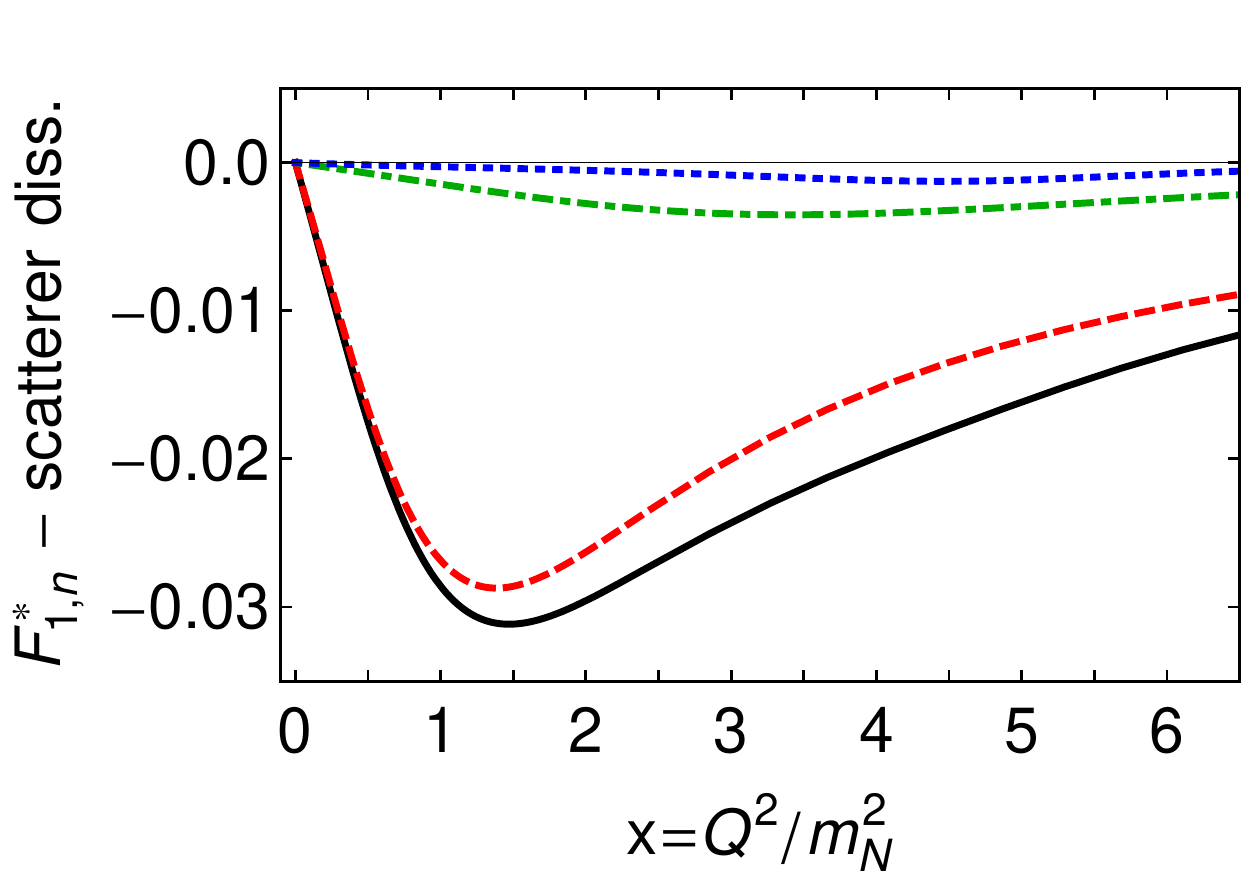}\vspace*{-1ex}
\end{tabular}
\end{center}
\caption{\label{figF1}
Computed Dirac transition form factor, $F_{1}^{\ast}$, for the charged reaction $\gamma^\ast\,p\to R^+$ (left panels) and the neutral reaction $\gamma^\ast\,n\to R^0$ (right panels): solid (black) curve in each panel.
Data, left panels:
circles (blue)~\cite{Aznauryan:2009mx},
and squares (purple)~\cite{Mokeev:2012vsa, Mokeev:2015lda}.
\emph{Upper panels} -- diquark breakdown: \emph{DD1} (dashed red), scalar diquark in both nucleon and Roper; \emph{DD2}
(dot-dashed green), pseudovector diquark in both nucleon and Roper; \emph{DD3} (dotted blue), scalar diquark in
nucleon, pseudovector diquark in Roper, and vice versa.
\emph{Lower panels} -- scatterer breakdown: \emph{DS1} (red dashed), photon strikes an uncorrelated dressed quark;
\emph{DS2} (dot-dashed green), photon strikes a diquark; and \emph{DS3} (dotted blue), diquark breakup contributions,
including photon striking exchanged dressed-quark.
}
\end{figure*}

Our predictions for the $\gamma^\ast N \to R$ Dirac transition form factors are drawn in Fig.\,\ref{figF1}.  They must vanish at $x=0$ owing to orthogonality between the nucleon and its radial excitation.
Plainly, this component of the charged transition proceeds primarily through a photon striking a bystander dressed-quark that is partnered by $[ud]$, with lesser but non-negligible contributions from all other processes.  In exhibiting these features, $F_{1,p}^{\ast}$ shows qualitative similarities to the elastic Dirac form factors depicted in Fig.\,\ref{elasticdissectF1}.
The neutral transition, too, proceeds primarily through a photon striking a bystander dressed-quark that is partnered by $[ud]$.  However, comparisons with the $R^0$ elastic Dirac form factor, left panels in Fig.\,\ref{elasticdissectF12neutral}, are complicated by the fact that charge neutrality enforces $F_1^{R^0}(0)=0$, so that all terms need only sum to zero at the origin, whereas state orthogonality ensures $F_{1,n}^\ast(0)=0$, in which case each contribution must vanish separately.

Regarding comparison with experiment, $F_{1,p}^\ast(x)$ agrees quantitatively in magnitude and trend with the data on $x\gtrsim 2$, an outcome which owes fundamentally to the QCD-derived momentum-dependence of the propagators and vertices employed in solving the bound-state and scattering problems.
The mismatch on $x\lesssim 2$ between data and the prediction is also revealing.  As we have emphasized, our calculation yields only those form factor contributions generated by a rigorously-defined dressed-quark core whereas meson-cloud contributions are expected to be important on $x\lesssim 2$.  Thus, the difference between the prediction and data may plausibly be attributed to MB\,FSIs, as described in Sec.\,5 of Ref.\,\cite{Roberts:2016dnb}.  (See also Ref.\,\cite{Mokeev:2015lda, Kamano:2018sfb}.)

The associated light-front-transverse transition charge-density is \cite{Tiator:2008kd}:
\begin{align}
\label{eqrhob}
\rho^{pR}(|\vec{b}|)
& := \int \frac{d^2 \vec{q}_\perp }{(2\pi)^2} \,{\rm e}^{i \vec{q}_\perp \cdot \vec{b}} F_1^\ast(|\vec{q}_\perp|^2)\,,
%
\end{align}
where $F_1^\ast$ is interpreted in a frame defined by $Q=(\vec{q}_\perp=(Q_1,Q_2),Q_3=0,Q_4=0)$. Plainly, $Q^2 = |\vec{q}_\perp|^2$.  Defined in this way, $\rho^{pR}(|\vec{b}|)$ has a straightforward quantum mechanical interpretation \cite{Miller:2007uy}.  A computation of $\rho^{pR}(|\vec{b}|)$ may be found elsewhere \cite{Roberts:2018hpf}, along with a related analysis of the impact of MB\,FSIs: see Eq.\,(19) therein and the related discussion.  Consistent with their role in reducing the nucleon and Roper quark-core masses, MB\,FSIs introduce significant attraction, working to screen the long negative tail of the quark-core contribution to $\rho^{pR}(|\vec{b}|)$, arising from pseudovector diquarks, and thereby compressing the transition domain in transverse space.

One must naturally ask whether similar remarks hold true for the neutral transition.  $F_{1,n}^\ast(x)$ is uniformly small and in such circumstances it is possible that MB\,FSIs are important on a larger $Q^2$-domain.  This appears to be the case, \emph{e.g}.\ with the electric quadrupole form factor in the $\gamma^\ast N \to \Delta$ transition \cite{Segovia:2014aza}.  On the other hand, $F_{1,n}^\ast(x)$ is similar in size to $F_1^{n}$, for which our framework provides a good description.  We judge, therefore, that our predictions for $F_{1,n}^\ast(x)$ and $F_{1,p}^\ast(x)$ should be reliable on comparable domains.

\begin{figure*}[!t]
\begin{center}
\begin{tabular}{cc}
\includegraphics[clip,width=0.43\linewidth]{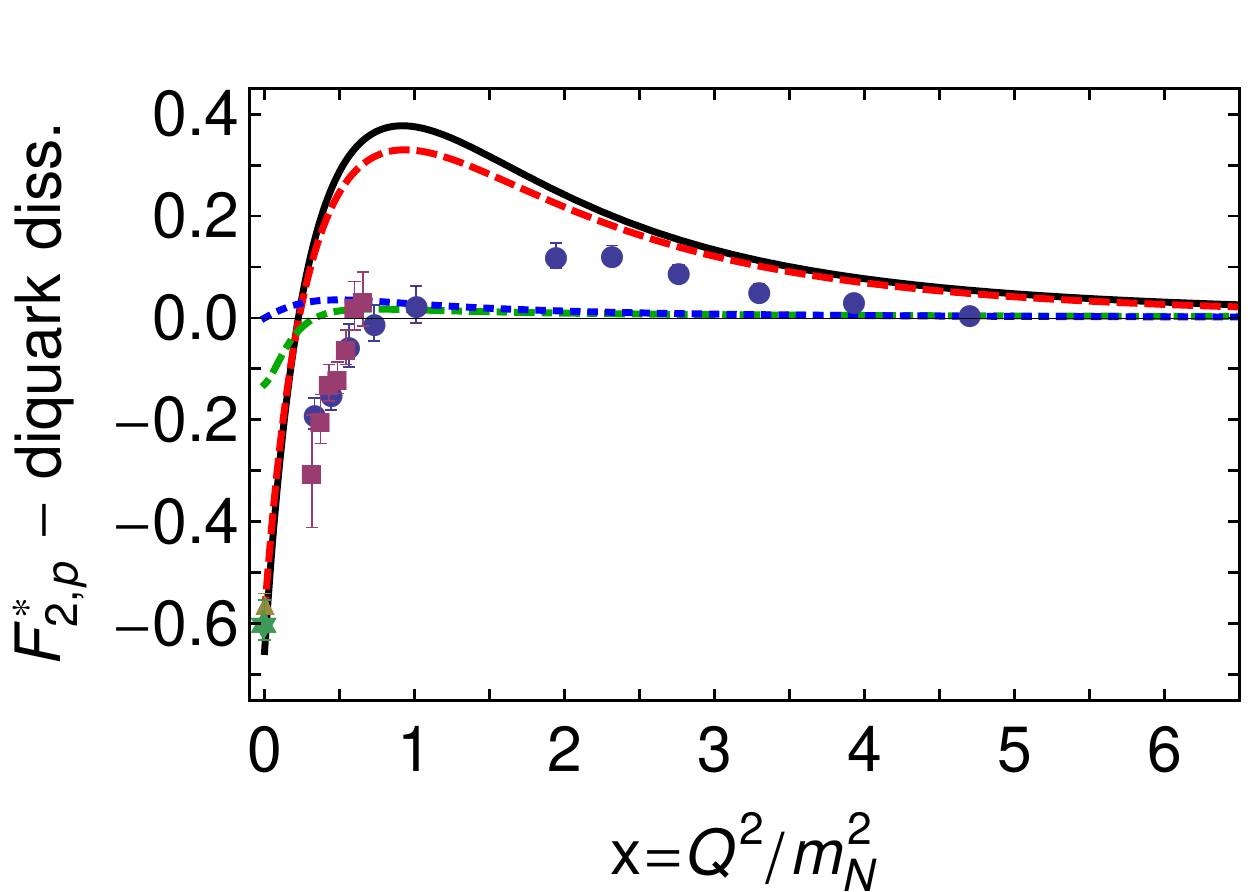}\hspace*{2ex } &
\includegraphics[clip,width=0.43\linewidth, height=0.235\textheight]{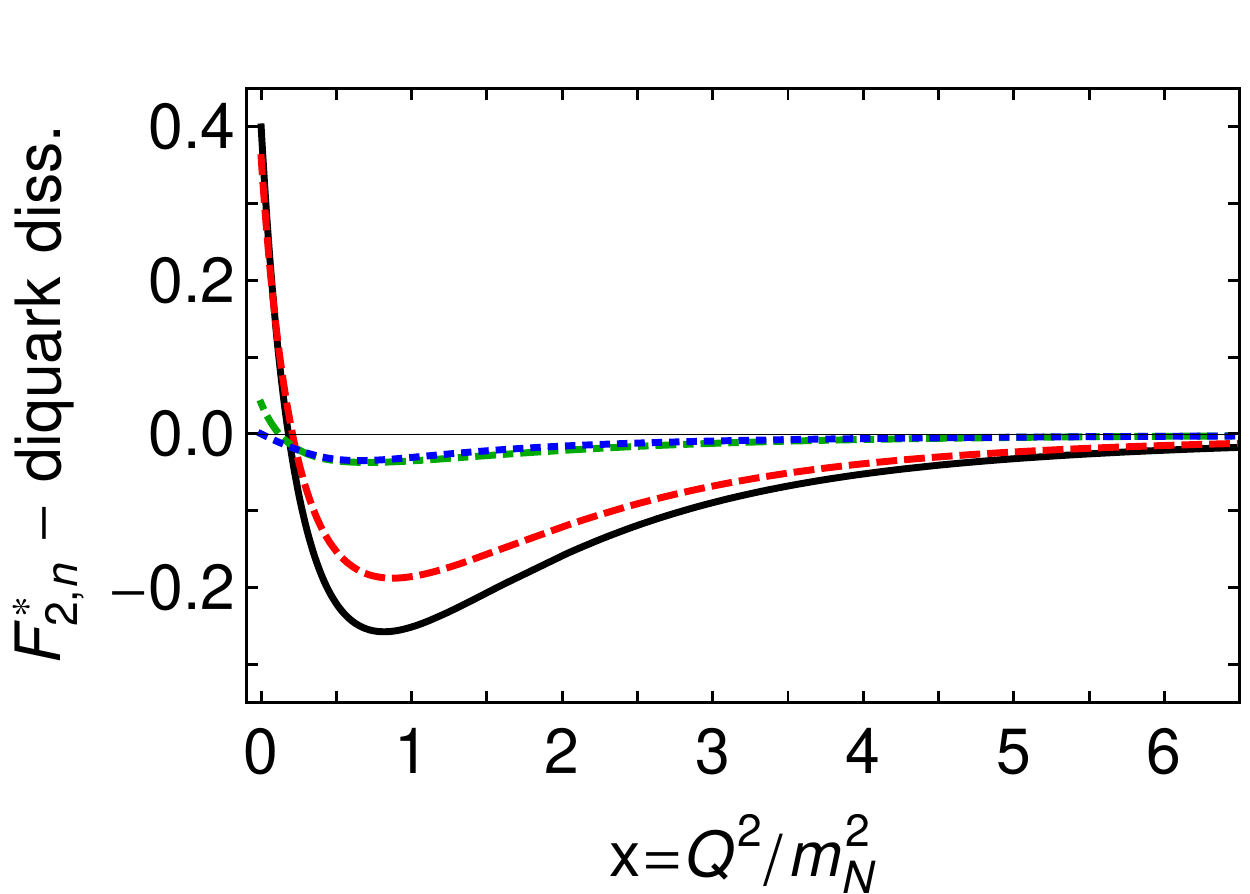}\vspace*{-0ex}
\end{tabular}
\begin{tabular}{cc}
\includegraphics[clip,width=0.43\linewidth]{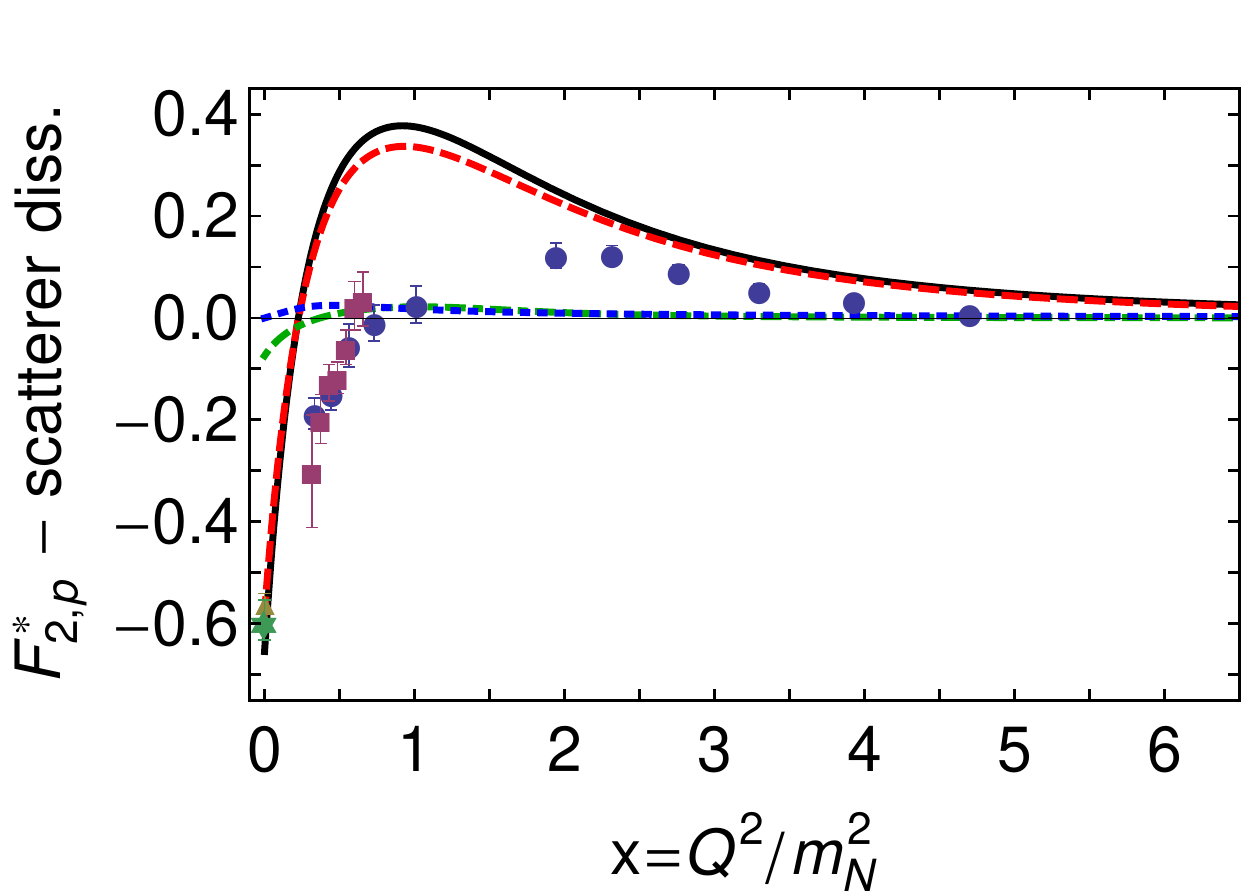}\hspace*{2ex } &
\includegraphics[clip,width=0.43\linewidth, height=0.235\textheight]{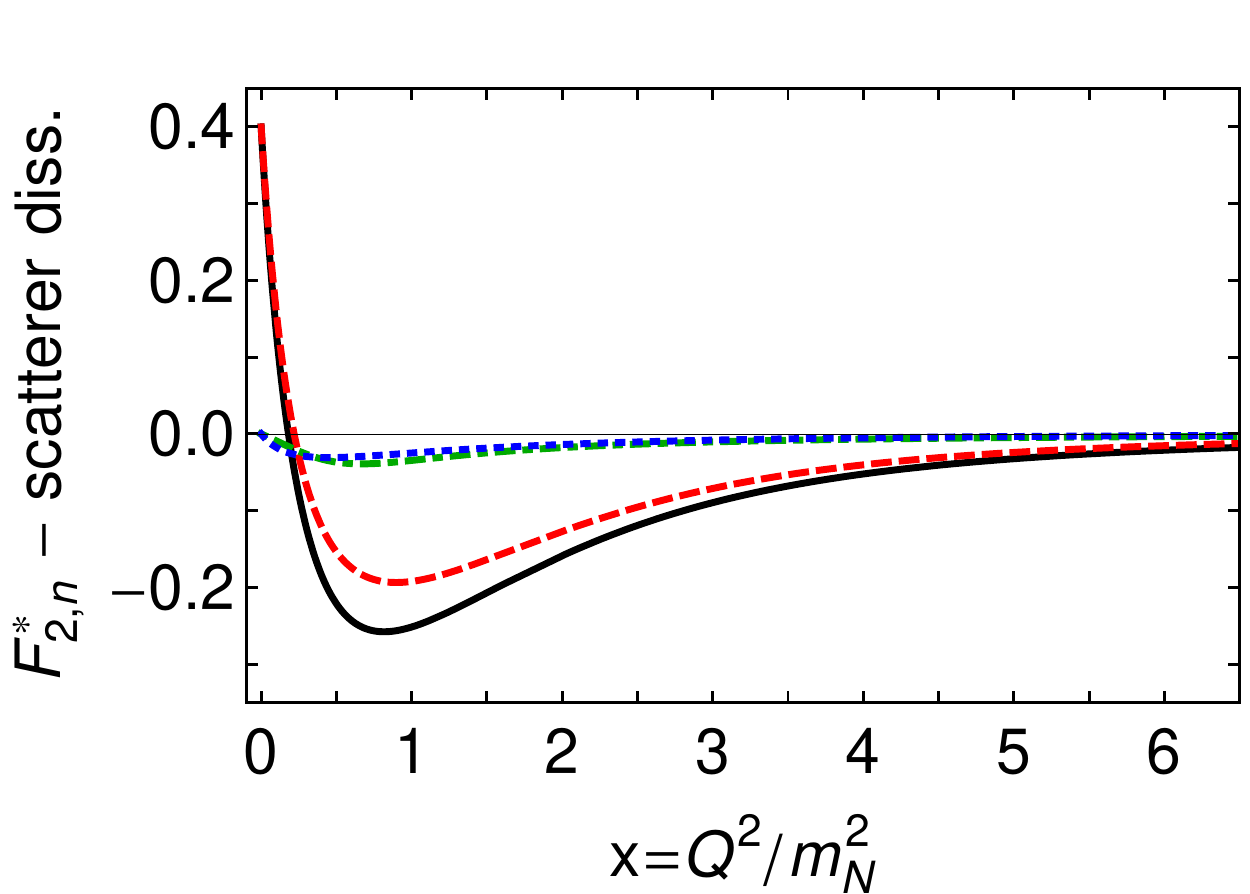}\vspace*{-1ex}
\end{tabular}
\end{center}
\caption{\label{figF2}
Computed Pauli transition form factor, $F_{2}^{\ast}$, for the charged reaction $\gamma^\ast\,p\to R^+$ (left panels) and the neutral reaction $\gamma^\ast\,n\to R^0$ (right panels): solid (black) curve in each panel.
Data:
circles (blue)~\cite{Aznauryan:2009mx},
squares (purple)~\cite{Mokeev:2012vsa, Mokeev:2015lda},
triangle (gold)~\cite{Dugger:2009pn},
and star (green)~\cite{Tanabashi:2018oca}.
\emph{Upper panels} -- diquark breakdown: \emph{DD1} (dashed red), scalar diquark in both nucleon and Roper; \emph{DD2}
(dot-dashed green), pseudovector diquark in both nucleon and Roper; \emph{DD3} (dotted blue), scalar diquark in nucleon, pseudovector diquark in Roper, and vice versa.
\emph{Lower panels} -- scatterer breakdown: \emph{DS1} (red dashed), photon strikes an uncorrelated dressed quark;
\emph{DS2} (dot-dashed green), photon strikes a diquark; and \emph{DS3} (dotted blue), diquark breakup contributions,
including photon striking exchanged dressed-quark.
}
\end{figure*}

Corresponding predictions for the $\gamma^\ast N \to R$ Pauli transition form factors are drawn in Fig.\,\ref{figF2}.  They are all nonzero at $x=0$ and each possesses a zero crossing at roughly the same location, \emph{viz}.\, $x\approx 0.2$.  Notably, as with $F_2^{p,R^+}$ and $F_2^{n,R^0}$ in Fig.\,\ref{elastic}, $F_{2,p}^\ast$ and $F_{2,n}^\ast$ are similar in magnitude and $Q^2$-dependence.
In particular, the value of $F_{2,p}^\ast(0)/F_{2,n}^\ast(0) \approx -3/2$ is consistent with available data \cite{Tanabashi:2018oca}.

The remarks above concerning MB\,FSIs also apply to $F_2^\ast$; and, importantly, although they affect its precise location, the existence of a zero in $F_{2}^{\ast}$ is not influenced by MB\,FSIs.  We are thus confident of our prediction for a zero in $F_2^{n,R^0}$.  This zero will be found near that of $F_2^{p,R^+}$ if MB\,FSIs are not too different between these channels; and there are good reasons to suppose they are comparable because the two reactions are isospin-exchange partners and isospin symmetry is a good approximation for strong interactions.

\section{Transitions at Larger $Q^2$}
\label{SecLargeQ2}
It is anticipated that the CLAS12 detector at JLab\,12 will deliver data on the Roper-resonance electroproduction form factors out to $Q^2 \approx 12\,m_N^2$ in both the charged and neutral channels.\footnote{In the latter case, although a deuteron must be used to provide the neutron target, there are indications that the quality of the cross-section data should be comparable to that for charged-Roper production off the free proton \cite{Tian:2018siq}.}  Here, therefore, we supply projections for all transition form factors on $x\in [0,12]$.

We must first remark that it is difficult to obtain reliable results for the form factors on $x> 6$ by direct calculation of all the contributions in Fig.\,\ref{vertexB} because Diagrams~3, 5, 6 are eight-dimensional integrals.  Monte-Carlo methods are required for their evaluation; but with any finite number of samples, such methods are imprecise when the answer is a small number, as is the case with form factors at large photon virtuality, and not all contributions are of the same sign.

We circumvent this difficulty by using the Schlessinger point method (SPM) \cite{Schlessinger:1966zz, PhysRev.167.1411, Tripolt:2016cya} to construct analytic approximations to each of our transition form factors on $x\in [0,6]$ and then defining the results on $x \in [6,12]$ via the analytic continuation of those approximations.

The SPM is based on the Pad\'e approximant.  It is able to accurately reconstruct a function in the complex plane within a radius of convergence specified by that one of the function's branch points which lies nearest to the real domain from which the sample points are drawn.  Moreover, owing to the procedure's discrete nature, the reconstruction may also provide a reasonable continuation on a larger domain, but this cannot be guaranteed and, hence, each case must be treated individually.

The following example provides a useful and relevant illustration. Suppose the function in question is a monopole form factor represented by a finite number, $N>0$, of points, each of which lies precisely on the curve.  Then using any single one of those points, the SPM will reproduce the monopole exactly.  If each of the points in the set has some numerical error, as is typical in the computation of form factors, then from any single point, the SPM will deliver an analytic approximation to the form factor.  Choosing a large number of single points at random and using the SPM with each point, then one obtains a collection of analytic approximations to the monopole whose spread measures the uncertainty inherent in the numerical calculation.  Each one of the approximations is of practically equal quality to the best least-squares fit.

For a realistic form factor, $F(Q^2)$, $\exists k\geq 0$ such that $(d/ d Q^2)^k F(Q^2)$ possesses at least one zero on $Q^2 > 0$.  In such cases one has a collection of $N$ results, each associated with the form factor, $F(Q^2)$, at a different value of $Q^2 \in [0, Q^2_{\rm max}]$, with $Q^2_{\rm max}$ being the upper bound of the domain upon which direct computations have been performed.  Each of the $N$ results will possess some similar level of precision.
From the set of $N$ results, one randomly chooses first one point, then two, etc., until reaching that minimal number of points, $M<N$, for which the analytic approximation produced by the SPM from any randomly chosen set of $M$ points typically delivers a valid fit to the output.
One then defines the extrapolation by randomly choosing a large number of $M$-point samples, determining the SPM approximation from each collection, applying any known physical constraints (such as continuity, known scaling behaviour, etc.) to eliminate those functions which are unacceptable, and then drawing the associated extrapolation curve for each surviving approximation.
This procedure generates a band of extrapolated curves whose collective reliability at any $Q^2>Q_{\rm max}^2$ is expressed by the width of the band at that point, which is itself determined by the precision of the original output on $Q^2 \leq Q^2_{\rm max}$.

\begin{figure}[!t]
\includegraphics[clip,width=0.86\linewidth]{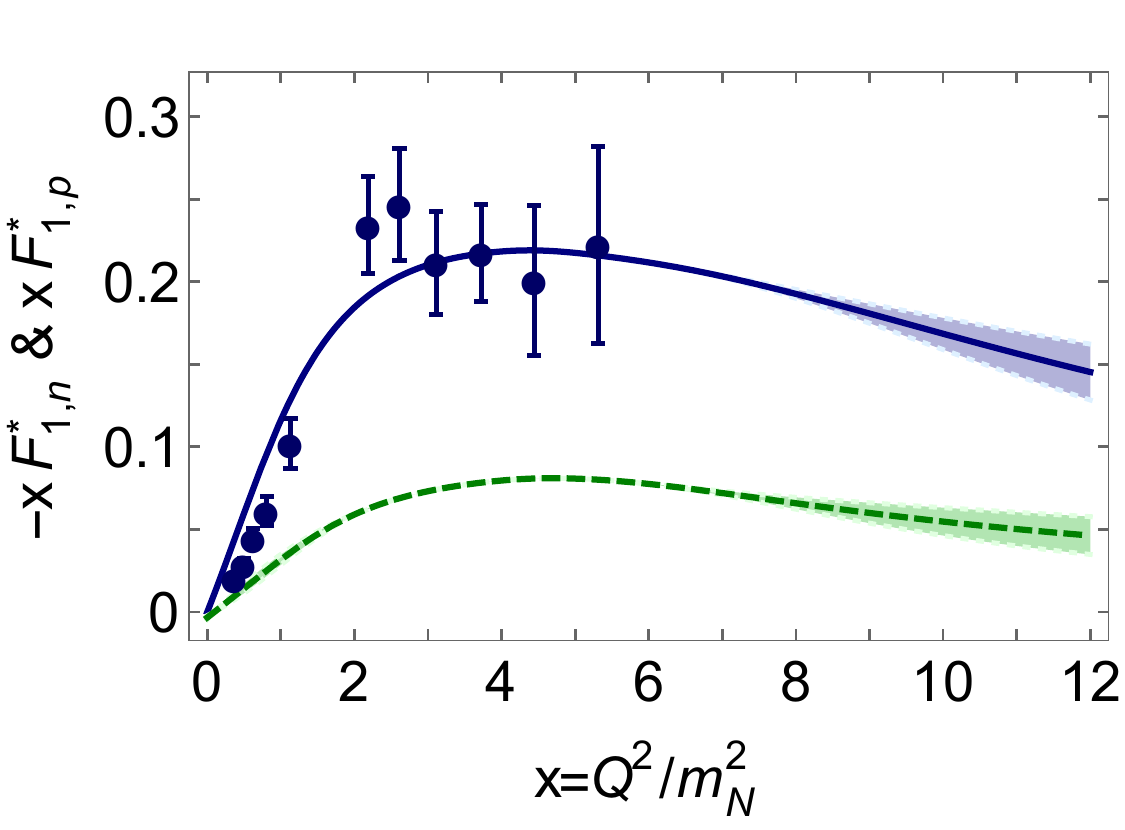}
\includegraphics[clip,width=0.86\linewidth]{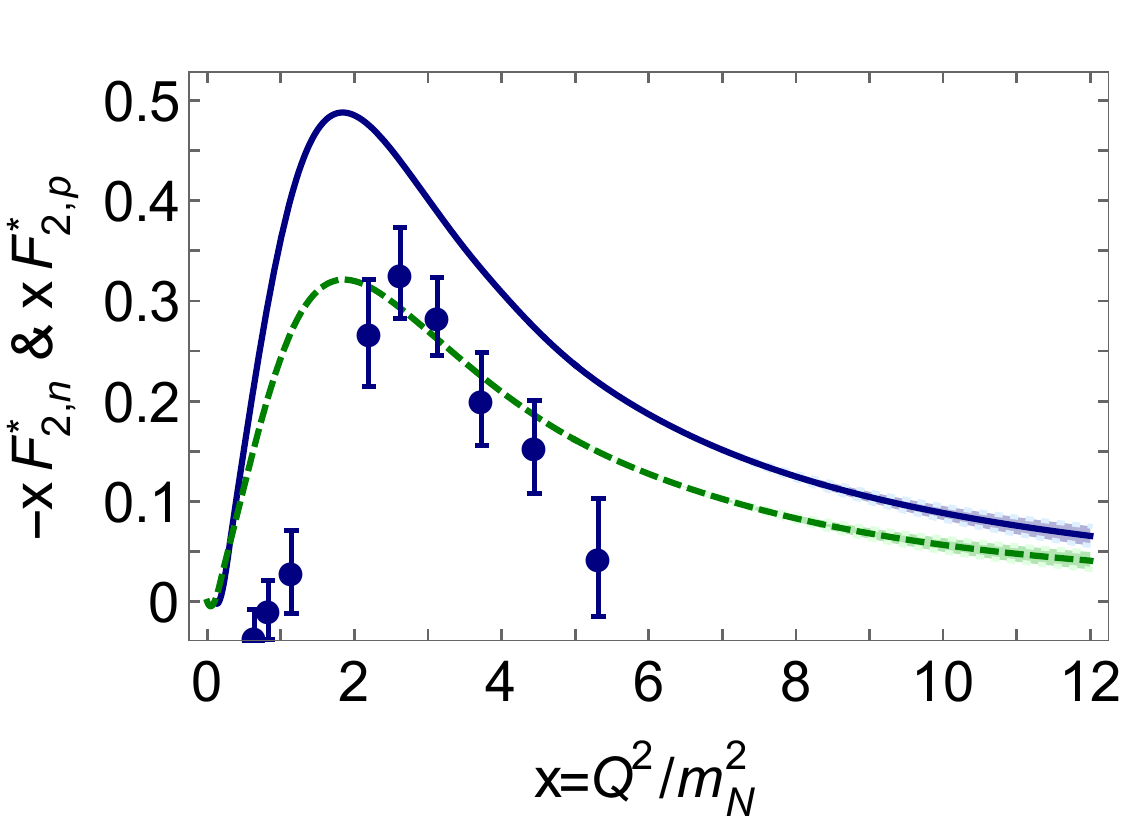}
\caption{\label{figLargeQ2}
Computed $x$-weighted Dirac (upper panel) and Pauli (lower panel) transition form factors for the reactions $\gamma^\ast\,p\to R^+$ (solid blue curves) and $\gamma^\ast\,n\to R^0$ (dashed green curves).  In all cases, the results on $x\in [6,12]$ are projections, obtained via extrapolation of analytic approximations to our results on $x\in [0,6]$: at each $x$, the width of the band associated with a given curve indicates our confidence in the extrapolated value.  (See text for details.)
Data in both panels are for the charged channel transitions, $F_{1,p}^\ast$ and $F_{2,p}^\ast$: circles (blue)~\cite{Aznauryan:2009mx}.  No data currently exist for the neutral channel.
}
\end{figure}

In Fig.\,\ref{figLargeQ2} we depict the $x$-weighted Dirac and Pauli transition form factors for the reactions $\gamma^\ast p \to R^{+}$, $\gamma^\ast n\to R^{0}$ on the domain $0<x<12$.  The results on $x>6$ were determined via the SPM, as described above: $M=12$ is satisfactory in all cases.  The precision of our projections can be exemplified by quoting the form factor values at the upper bound of the extrapolation domain, $x_{12}=12$:
\begin{subequations}
\begin{align}
 F^\ast_{1,p} & = 0.0121(14) \,, \;  & -F^\ast_{1,n} & = 0.0039(10) \,,\\
x_{12}F^\ast_{1,p} & = 0.145(17) \,, \; & -x_{12}F^\ast_{1,n}  &= 0.046(11) \,,
\end{align}
\end{subequations}
\vspace*{-7ex}

\begin{subequations}
\begin{align}
F^\ast_{2,p} & = 0.0055(8) \,, \; & -F^\ast_{2,n}  &= 0.0034(7) \,,\\
x_{12}F^\ast_{2,p} & = 0.066(10) \,, \; & -x_{12}F^\ast_{2,n}  &= 0.041(9) \,.
\end{align}
\end{subequations}

We choose to draw $x$-weighted results in order to accentuate, but not overmagnify, the larger-$x$ behaviour of the form factors.  On the domain depicted, there is no indication of the scaling behaviour expected of the transition form factors: $F^\ast_{1} \sim 1/x^2$, $F^\ast_2 \sim 1/x^3$.  Since each dressed-quark in the baryons must roughly share the impulse momentum, $Q$, we expect that such behaviour will only become evident on $x\gtrsim 20$.

\section{Flavour Separation}
\label{secFlavourSeparated}
If one supposes that $s$-quark contributions to $N\to R$ transitions are negligible, as they are in nucleon elastic form factors, and assumes isospin symmetry, as we do, then a flavour separation of the transition form factors is accomplished by combining results for the $\gamma^\ast p\to  R^+$ and $\gamma^\ast n\to R^0$ transitions:
\begin{equation}
\label{FlavourSep}
F_{i,u}^{\ast} = 2 F_{i,p}^{\ast} + F_{i,n}^{\ast}, \;
F_{i,d}^{\ast} = F_{i,p}^{\ast} + 2 F_{i,n}^{\ast},\; i=1,2\,.
\end{equation}
Our conventions are that $F_{1(2),u}^{\ast}$ and $F_{1(2),d}^{\ast}$ refer to the $u$- and $d$-quark contributions to the equivalent Dirac (Pauli) form factors of the $\gamma^\ast p\to R^+$ reaction, and the results are normalised such that the \emph{elastic} Dirac form factors of the proton and charged-Roper yield $F_{1u}(Q^2=0)=2$, $F_{1d}(Q^2=0)=1$, thereby ensuring that these functions count valence $u$- and $d$-quark content in the bound-states.

\begin{figure}[!t]
\includegraphics[clip,width=0.86\linewidth]{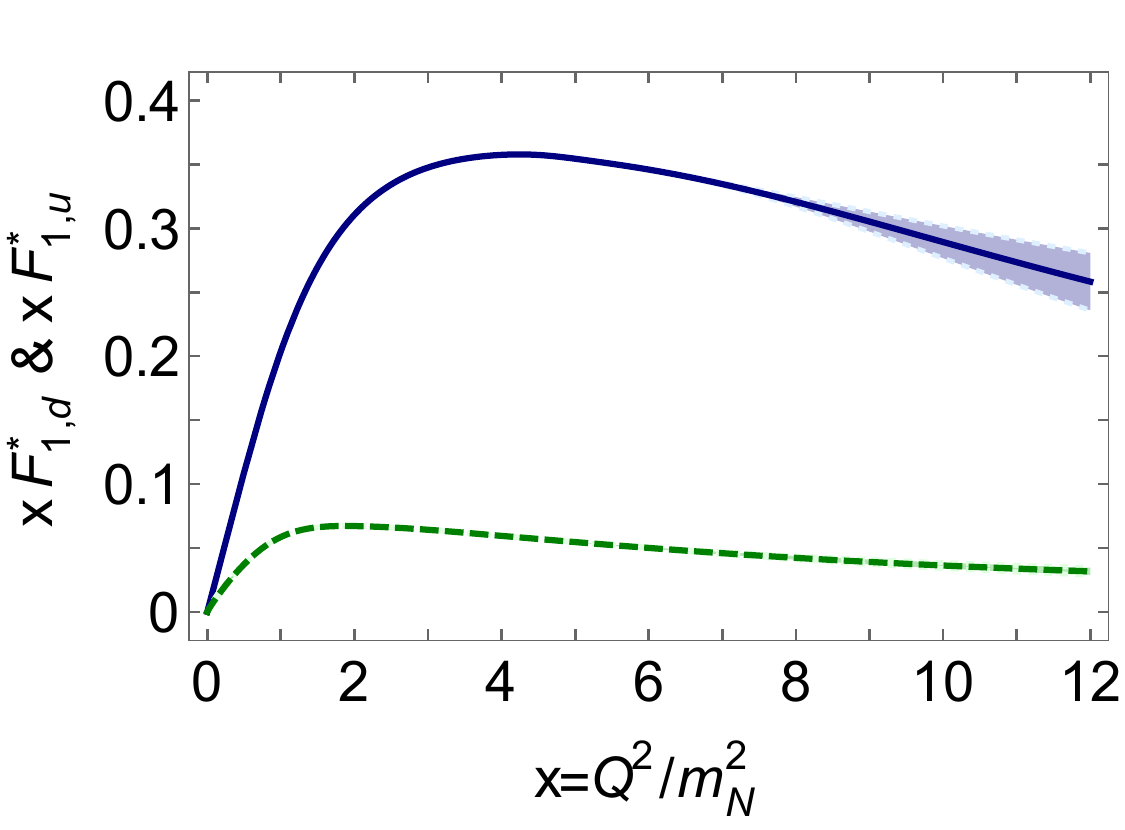}
\includegraphics[clip,width=0.86\linewidth]{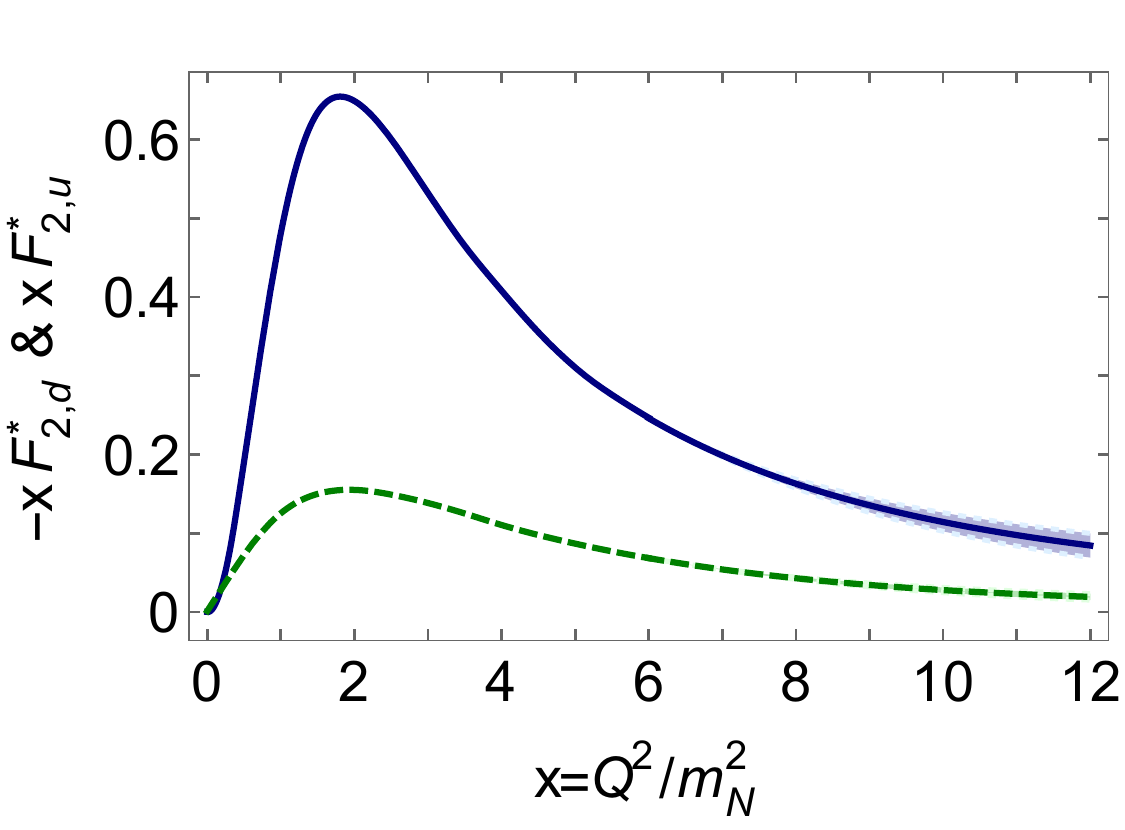}
\caption{\label{FlavourSeparated}
Flavour separation, $x$-weighted $\gamma^\ast\,p\to R^+$ transition form factors: $u$-quark, solid (blue); and $d$-quark, dashed (green).  \emph{Upper panel} -- Dirac transition form factor.  \emph{Lower panel} -- Pauli transition form factor.  As in Fig.\,\ref{figLargeQ2}, at each $x\geq 6$, the width of the band associated with a given curve indicates our confidence in the extrapolated value.  (For reference, $\kappa_u^\ast  = F_{2,u}^\ast(0) = -0.91$, $\kappa_d^\ast = F_{2,d}^\ast(0) = 0.14$.)
}
\end{figure}

Using the transition form factors reported in Sec.\,\ref{secEMtransition}, one obtains the flavour-separated results described in Ref.\,\cite{Segovia:2016zyc}.  In Fig.\,\ref{FlavourSeparated}, we reproduce those results and extend them to the domain $6<x<12$ using the SPM, as described in Sec.\,\ref{SecLargeQ2}.

The flavour-separated Dirac transition form factors depicted in the upper panel of Fig.\,\ref{FlavourSeparated} behave similarly to the analogous elastic form factors reported elsewhere \cite{Segovia:2014aza}: the $d$-quark contribution is markedly smaller than the $u$-quark contribution on the entire domain explored.  The noticeable difference, however, is the absence of a zero in $F_{1,d}^{\ast}$ on this domain.  Such a zero is a salient feature of both the analogous proton elastic Dirac form factor (located at $x\approx 7$) and the Roper elastic Dirac form factor ($F_{1,d}^{R^+}(x \approx 5)=0$).  Its absence in $F_{1,d}^{\ast}$ here owes to orthogonality of the initial and final state wave functions and the attendant redistribution of integration support in computation of the transition form factors.

The lower panel of Fig.\,\ref{FlavourSeparated} depicts the flavour-separated Pauli transition form factor.  Once again, the behaviour of the $u$- and $d$-quark contributions to the charged-Roper Pauli transition form factor are comparable with the kindred contributions to the elastic form factor.

As highlighted elsewhere \cite{Segovia:2016zyc}, an explanation for the pattern of behaviour in Fig.\,\ref{FlavourSeparated} is similar to that for the analogous proton elastic form factors \cite{Segovia:2015ufa} because the diquark content of the proton and its first radial excitation are almost identical.  In both systems, the dominant piece of the associated Faddeev wave functions is $\psi_0$, namely a $u$-quark in tandem with a $[ud]$ (scalar diquark) correlation, which produces $\approx 60$\% of each bound-state's normalisation.  If $\psi_0$ were the sole component in both the proton and charged-Roper, then $\gamma$--$d$-quark interactions would receive a $1/x$ suppression on $x>1$, because the $d$-quark is sequestered in a soft correlation, whereas a spectator $u$-quark is always available to participate in a hard interaction.  At large $x$, therefore, scalar diquark dominance leads one to expect $F^\ast_d \sim F^\ast_u/x$.  Naturally, the details of this $x$-dependence are influenced by the presence of pseudovector diquark correlations in the initial and final states, which guarantees that the singly-represented $d$-quark, too, can participate in a hard scattering event, but to a lesser extent.

The infrared behaviour of the flavour-separated $\gamma^\ast p \to R^+$ transition form factors owes to an intricate interference between the influences of orthogonality, which forces $F^\ast_{1,u}(x=0)=0=F^\ast_{1,d}(0)$, and quark-core and MB\,FSI contributions.  However, whilst the latter pair act in similar ways for both elastic and transition form factors, orthogonality is a fundamental difference between the two processes.  It is therefore likely to be the dominant effect at infrared momenta.

The information contained in Figs.\,\ref{figF1} -- \ref{FlavourSeparated} provides evidence in support of the notion that many features in the measured behaviour of $\gamma^\ast N \to R$ electromagnetic transition form factors are primarily driven by the presence of strong diquark correlations in the nucleon and its first radial excitation.  In our view, inclusion of a ``meson cloud'' cannot qualitatively affect the salient features of these transition form factors, any more than it does the analogous nucleon elastic form factors.

\section{Epilogue}
\label{secEpilogue}
It has long been argued that the phenomenon of dynamical chiral symmetry breaking in the Standard Model entails that baryon wave functions possess strong nonpointlike quark-quark (diquark) correlations, whose presence has many observable consequences.  We therefore used a quark-diquark approximation to the Poincar\'e-covariant three-body bound-state problem to compute all form factors relevant to the $\gamma^\ast N \to R$ transitions.  In doing so we completed a unification of $\gamma^\ast N \to R$ transition form factors with nucleon elastic and $\gamma^\ast N \to \Delta(1232)$ transition form factors: both scalar and pseudovector diquarks are essential for a description of existing data in all these cases, but correlations in other diquark channels can be neglected.

Focusing on $\gamma^\ast N \to R$, precise measurements in the charged channel already exist \cite{Aznauryan:2009mx, Aznauryan:2012ec, Aznauryan:2012ba, Park:2014yea, Isupov:2017lnd, Fedotov:2018oan}, novel experiments are approved at JLab\,12 and elsewhere, and others are either planned or under consideration as part of an international effort to measure transition electrocouplings of all prominent nucleon resonances \cite{Aznauryan:2012ba, E12-09-003, E12-06-108A, Mokeev:2018zxt, Carman:2018fsn, Cole:2018faq, Ramstein:2018wkk}.  Hence, it is likely that our predictions, including those in the neutral channel, will be tested in the foreseeable future.

Such experiments have the potential to deliver empirical information that would address a wide range of issues, including, \emph{e.g}.: is the expression of DCSB the same in each baryon; and are quark-quark correlations an essential element in the structure of all baryons?  Modern calculations answer ``yes'' to both questions for the nucleon, $\Delta$-baryon and Roper resonance.  However, these three systems are merely a small collection of closely-related positive-parity baryons; and, hence, consistency with available data may be seen as suggestive but not conclusive.  This is especially true given emerging evidence which indicates that pseudoscalar and vector diquark correlations also play a material role in low-lying negative-parity baryons \cite{Eichmann:2016hgl, Lu:2017cln, Chen:2017pse}, excited states of the $\Delta$-baryon possess unexpectedly complicated wave functions \cite{Qin:2018dqp}, and very little is known about the Poincar\'e-covariant wave functions of $I=1/2$, $J=3/2$ baryons.  These systems are the focus of forthcoming analyses using the methods described herein.  The lesson of the Roper resonance shows that only precise measurements of the associated transition form factors on $Q^2 \gtrsim 2\,m_N^2$, \emph{i.e}.\ beyond the meson-cloud domain, will be capable of validating whatever global picture of baryon structure emerges from Poincar\'e-covariant studies of the continuum bound-state problem.

\medskip

\centerline{\textbf{ACKNOWLEDGMENTS}}

\smallskip

We are grateful for constructive comments and encouragement from V.\,Burkert, L.~Chang, P.~Cole, \mbox{Z.-F.~Cui}, R.~Gothe, G.~Krein, V.~Mokeev, B.~Ramstein, S.\,M.~Schmidt, F.~Wang and \mbox{H.-S.~Zong};
and for the hospitality and support of the University of Huelva, Huelva - Spain, and the University of Pablo de Olavide, Seville - Spain, during the ``4th Workshop on Nonperturbative QCD'' at the University of Pablo de Olavide, 6-9 November 2018.
Work supported by:
Funda\c{c}\~ao de Amparo \`a Pesquisa do Estado de S\~ao Paulo - FAPESP Grant No.\,2015/21550-4;
National Natural Science Foundation of China, under grant no.\ 11805097;
Jiangsu Province Natural Science Foundation, under grant no.\ BK20180323;
U.S.\ Department of Energy, Office of Science, Office of Nuclear Physics, under contract no.\,DE-AC02-06CH11357;
Chinese Ministry of Education, under the \emph{International Distinguished Professor} programme;
Jiagnsu Province \emph{Hundred Talents Plan for Professionals};
Spanish MEyC, under grant no.\ FPA2017-86380-P;
European Union's Horizon 2020 research and innovation programme under the Marie Sk\l{}odowska-Curie grant agreement no.\,665919;
and the Spanish MINECO's Juan de la Cierva-Incorporaci\'on programme with grant agreement no.\,IJCI-2016-30028.
%



\end{document}